\documentclass[twocolumn]{aastex631}

\newcommand{\Msun}{ \mbox{M$_\odot$}}
\newcommand{\Mjup}{\mbox{M$_{\rm Jup}$}}
\newcommand{\Teff}{T$_{\rm{eff}}$}
\usepackage{amssymb,gensymb,times,graphicx,morefloats,color}
\usepackage{amsmath}
\usepackage{graphics,graphicx,url,verbatim,xspace}

\usepackage{changepage}
\usepackage{mathtools} 
\usepackage{comment}
\usepackage{gensymb}
\usepackage{hyperref}

\newcommand\h{$^{\mathrm{h}}$}
\newcommand\m{$^{\mathrm{m}}$}
\newcommand\s{$^{\mathrm{s}}_{.}$}

\begin{document}
\defcitealias{Ren2020AFGKSample}{R20}
\defcitealias{Holberg2013}{H13}
\defcitealias{Willems2004DetechedWDMS}{W04}
\defcitealias{koesterWhiteDwarfSpectra2010}{K10}
\defcitealias{Bergeron1995CoolingModels}{B95}

\title{Five New Sirius-Like White Dwarf + Main Sequence Star Systems with MagAO-X}

\shorttitle{New SLS with MagAO-X}
\shortauthors{Pearce et al.}

\correspondingauthor{Logan A. Pearce}
\email{lapearce@umich.edu}

\author[0000-0003-3904-7378]{Logan A. Pearce}
\affil{Department of Astronomy, University of Michigan, Ann Arbor, MI 48109, USA}

\author[0000-0002-2346-3441]{Jared R. Males}
\affil{Steward Observatory, University of Arizona, Tucson, AZ 85721, USA}

\author[0000-0001-5130-9153]{Sebastiaan Y. Haffert}
\affil{Leiden Observatory, Leiden University, PO Box 9513, 2300 RA Leiden, The Netherlands}
\affil{Steward Observatory, University of Arizona, Tucson, AZ 85721, USA}

\author[0000-0002-2167-8246]{Laird M. Close}
\affil{Steward Observatory, University of Arizona, Tucson, AZ 85721, USA}

\author[0000-0003-1905-9443]{Joseph D. Long}
\affil{Center for Computational Astrophysics, Flatiron Institute, 162 5th Avenue, New York, New York}

\author[0000-0003-0843-5140]{Eden A. McEwen}
\affil{James C. Wyant College of Optical Sciences, University of Arizona, 1630 E University Blvd, Tucson, AZ, 85719, USA}

\author[0000-0002-4934-3042]{Joshua Liberman}
\affil{James C. Wyant College of Optical Sciences, University of Arizona, 1630 E University Blvd, Tucson, AZ, 85719, USA}

\author[0000-0003-3253-2952]{Maggie Y. Kautz}
\affil{Steward Observatory, University of Arizona, Tucson, AZ 85721, USA}
\affil{James C. Wyant College of Optical Sciences, University of Arizona, 1630 E University Blvd, Tucson, AZ, 85719, USA}

\author{Jay K. Kueny}
\affil{James C. Wyant College of Optical Sciences, University of Arizona, 1630 E University Blvd, Tucson, AZ, 85719, USA}

\author[0000-0001-6654-7859]{Alycia J. Weinberger}
\affil{Earth \& Planets Laboratory, Carnegie Science, 5241 Broad Branch Road NW, Washington, DC}

\author[0000-0002-8110-7226]{Jialin Li}
\affil{Steward Observatory, University of Arizona, Tucson, AZ 85721, USA}

\author[0009-0006-4370-822X]{Elena Tonucci}
\affil{Leiden Observatory, Leiden University, PO Box 9513, 2300 RA Leiden, The Netherlands}

\author[0009-0002-9752-2114]{Katie Twitchell}
\affil{James C. Wyant College of Optical Sciences, University of Arizona, 1630 E University Blvd, Tucson, AZ, 85719, USA}

\author{Avalon L. McLeod}
\affil{Draper Laboratory, 555 Technology Square, Cambridge, Massachusetts}

\author{Warren B. Foster}
\affil{Steward Observatory, University of Arizona, Tucson, AZ 85721, USA}
\affil{James C. Wyant College of Optical Sciences, University of Arizona, 1630 E University Blvd, Tucson, AZ, 85719, USA}

\author{Olivier Guyon}
\affil{Steward Observatory, University of Arizona, Tucson, AZ 85721, USA}
\affil{James C. Wyant College of Optical Sciences, University of Arizona, 1630 E University Blvd, Tucson, AZ, 85719, USA}
\affil{National Astronomical Observatory of Japan, Subaru Telescope, National Institutes of
Natural Sciences, Hilo, HI 96720, USA}
\affil{Astrobiology Center, National Institutes of Natural Sciences, 2-21-1 Osawa, Mitaka, Tokyo, JAPAN}

\author{Alexander D. Hedglen}
\affil{Northrop Grumman Corporation, 600 South Hicks Road, Rolling Meadows, Illinois}

\author{Kyle Van Gorkom}
\affil{Steward Observatory, University of Arizona, Tucson, AZ 85721, USA}

\author{Jennifer Lumbres}
\affil{Steward Observatory, University of Arizona, Tucson, AZ 85721, USA}
\affil{James C. Wyant College of Optical Sciences, University of Arizona, 1630 E University Blvd, Tucson, AZ, 85719, USA}

\author{Lauren Schatz}
\affil{Kirtland Air Force Base, Air Force Research Laboratory, Albuquerque, NM, USA}

\author{Victor Gasho}
\affil{Steward Observatory, University of Arizona, Tucson, AZ 85721, USA}

\author{Katie M. Morzinski}
\affil{Steward Observatory, University of Arizona, Tucson, AZ 85721, USA}

\author{Phil M. Hinz}
\affil{UC Santa Cruz, 1156 High St, Santa Cruz CA 95064, USA}

\begin{abstract}
Most known white dwarfs in multiple systems with main sequence stars have been discovered with M-type companions, because the white dwarf causes detectable UV excess and bluer colors than expected from a single M star. Surveys have shown that the number of white dwarfs in Sirius-like systems within 100 pc of the Sun is lower than expected, suggesting that white dwarfs are being missed in the glare of their main sequence companions.  In this work we have leveraged the angular resolution and high-contrast capabilities, as well as optimization for visible wavelengths, of the extreme adaptive optics instrument MagAO-X to detect new white dwarf companions to AFGK stars. We present the first results of our survey with the extreme AO instrument MagAO-X, called the Pup Search, of 18 targets with seven new candidate companions, five of which are confirmed to be white dwarfs. We discuss the new detections in the context of previous surveys and other detection metric sensitivities and show that we are sensitive to a region not probed by other surveys. Finally we discuss the future of the Pup Search in light of developing technologies.
\end{abstract}


\section{Introduction} \label{sec:intro}

The majority of known white dwarfs (WDs) in multiple systems with main sequence stars (MS) are WD companions to M-dwarf stars. WD companions are easier to find around M dwarfs where the WD-M dwarf contrasts are low and the WD dominates the blue end of the system spectra energy distribution. For earlier type stars (AFGK spectral types), the WD companion signal can be drowned out by the brighter, bluer (compared to M stars) MS star signal. WD companions to such so-called Sirius-like Systems (SLS; \citealt[][ hereafter \citetalias{Holberg2013}]{Holberg2013}) are rare compared to evolutionary predictions (50-60\% of BAF stars are binaries yet only 32$\pm$8\% are observed, \citealt{Holberg2009WDBinaries, Ferrario2012}), indicating that many WD companions are likely being missed in the glare of their brighter hosts, particularly if they are close enough to be unresolvable in imaging. Previous SLS surveys \citep[e.g. ][]{zuckermanOccurrenceWideorbitPlanets2014, Noor2024WDPollutionInWideBinaries} were typically limited to separations~$\gtrsim 1$''. Surveys targeting more massive AFGK type stars at closer separations will continue to fill in the SLS population parameter space.

The new generation of ground-based, extreme adaptive optics (ExAO) high-contrast imaging instrumentation is well suited to probe tighter SLS systems than previous surveys. ExAO instruments are designed to work at high contrasts of 10$^{-6}$ and better and close angular resolutions of 10 -- 100's of mas for direct detection of exoplanets.  Close WD companions will have much lower contrasts of 10$^{-3}$ -- 10$^{-4}$ in visible wavelengths, allowing relatively simple detection. Additionally, the ExAO instruments MagAO-X \citep{Males2022MagAOX} and SCExAO/VAMPIRES \citep{Norris2015SCExAOVampires, Evans2022VampiresCoronagraph}, unlike other AO systems optimized for Near/Mid-infrared wavelengths, are optimized for visible wavelengths (500-1000~nm), where WD-MS flux contrasts will be lower, making them ideal instruments for this science case.

We are leveraging the power of ExAO instruments towards these problems with a survey called The ExAO Pup Search: The extreme AO non-interacting white dwarf-main sequence binary system survey.\footnote{The Pup Search name is a reference to the first known wide White Dwarf - Main Sequence system, Sirius AB, discovered in 1844 by Friedrich Bessel when he observed changes in the proper motion of Sirius \citep{Bessel1844VariationsProperMotions}, first observed by Alvin Graham Clark \citep{Flammarion1877CompanionofSirius}, and confirmed as the second known WD via its spectrum obtained by Walter Adams \citep{Adams1915SpectrumOfCompanionSirius}. Since Sirius A is the ``Dog Star", Sirius B was nicknamed ``The Pup"} \added{The ExAO Pup Search targets a set of AFGK stars with UV excess that have been vetted with radial velocities to exclude close companions, delivering a sample with a high likelihood of hosting an accessible WD companion.} Newly discovered SLS in a regime not probed by other surveys will provide a population for continued study of main sequence companion influence on pollution.

We observed an initial set of Pup Search targets in 2022 and 2024 and detected five new WDMS star system candidates, one new stellar binary, and one candidate unresolved WD+Md companion. \deleted{In this work we present the initial detection and photometry of new WD candidate companions.}  In Section \ref{sec:observations} we describe the Pup Search target selection, observations, and data reduction.  In Section \ref{sec:results} we report new confirmed and candidate signal detections, non-detections, and contrast curves. In Section 4 we discuss implications of these new detections in the wider SLS picture.

\section{Observations}\label{sec:observations}

In this section we describe selection of targets with a high likelihood of hosting a WD companion accessible to MagAO-X, the observations of 18 targets reported in this work, and data reduction.

\subsection{Target Selection}\label{sec: targets}

Our Pup Search targets are drawn from the catalog produced by the White Dwarf Binary Pathways Survey \citep{Parsons2016Pathways1} in \citealt[][]{Ren2020AFGKSample} (hereafter \citetalias{Ren2020AFGKSample}). The White Dwarf Binary Pathways Survey is interested in post-common envelope WD + main sequence star systems (PCEB). \citetalias{Ren2020AFGKSample} presented radial velocities of 275 WD+AFGK star candidates identified from TGAS (Tycho-Gaia Astrometric Solution, \citealt{Michalik2015TGAS}), \textsl{Gaia} DR2 \citep{GaiaCollaboration2016, GaiaDR22018}, and GALEX \citep{Bianchi2017GALEX} UV excess from the \citet{Parsons2016Pathways1} sample. They measured $\ge$2 RV observations for 151 systems, and selected 23 candidate systems most likely to be close PCEBs based on RV signal, which we excluded from our target list. This provides a well-vetted catalog of 128 candidate WD+AFGK systems with separations potentially accessible to ExAO.  From those 128 we selected 84 which were not identified as spectroscopic binaries in Simbad \citep{Wenger2000SIMBAD} and for which the MS star is bright enough for ExAO (\textsl{Gaia} G magnitude $<$ 11) and at declinations accessible to MagAO-X (-80$^{\circ}$ -- +20$^{\circ}$). \added{The catalog identifies stars by TYC number, so we report TYC number in addition to other identifiers in this work.}

\begin{figure}
\centering
\includegraphics[width=0.45\textwidth]{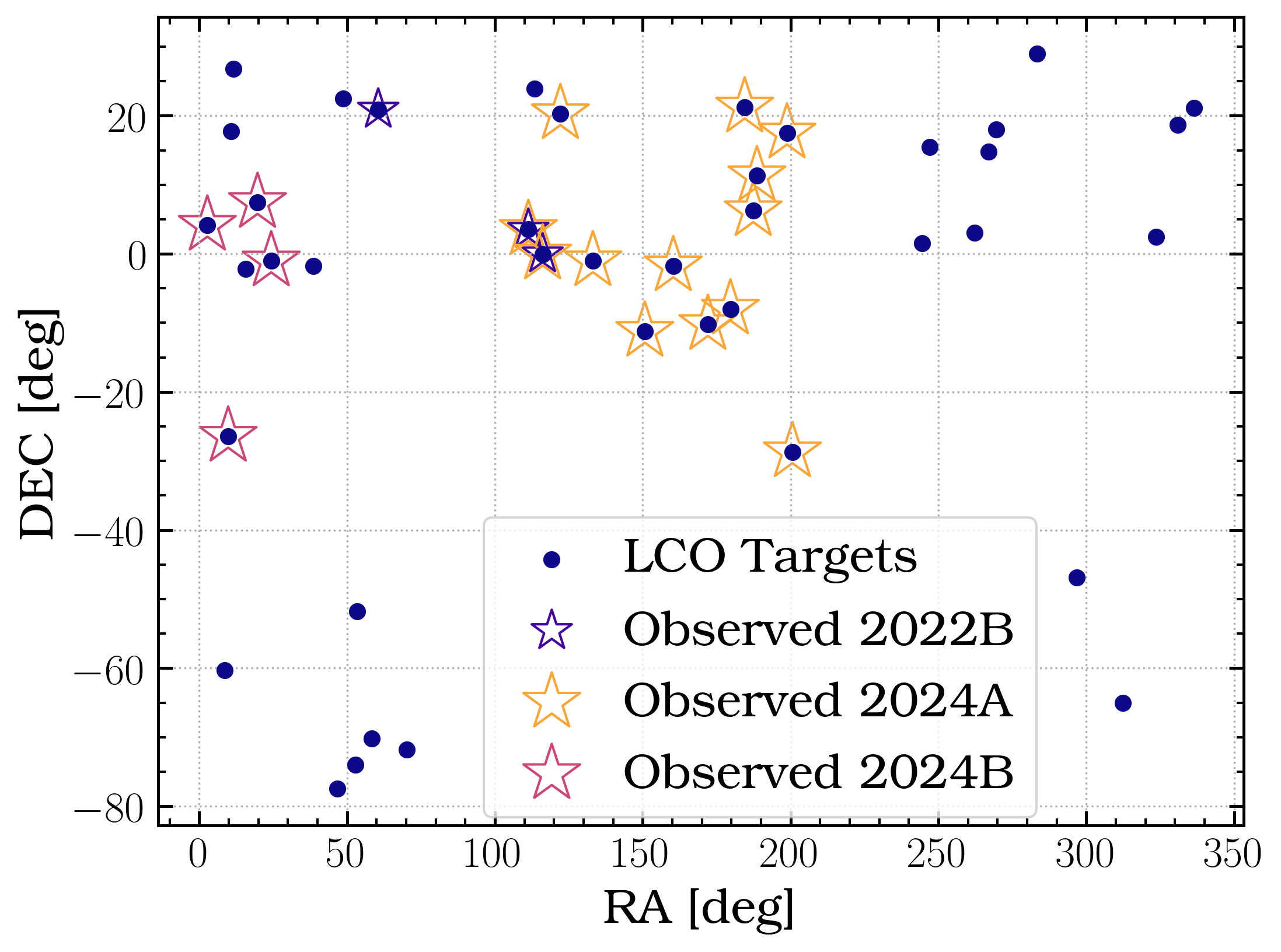}
\includegraphics[width=0.45\textwidth]{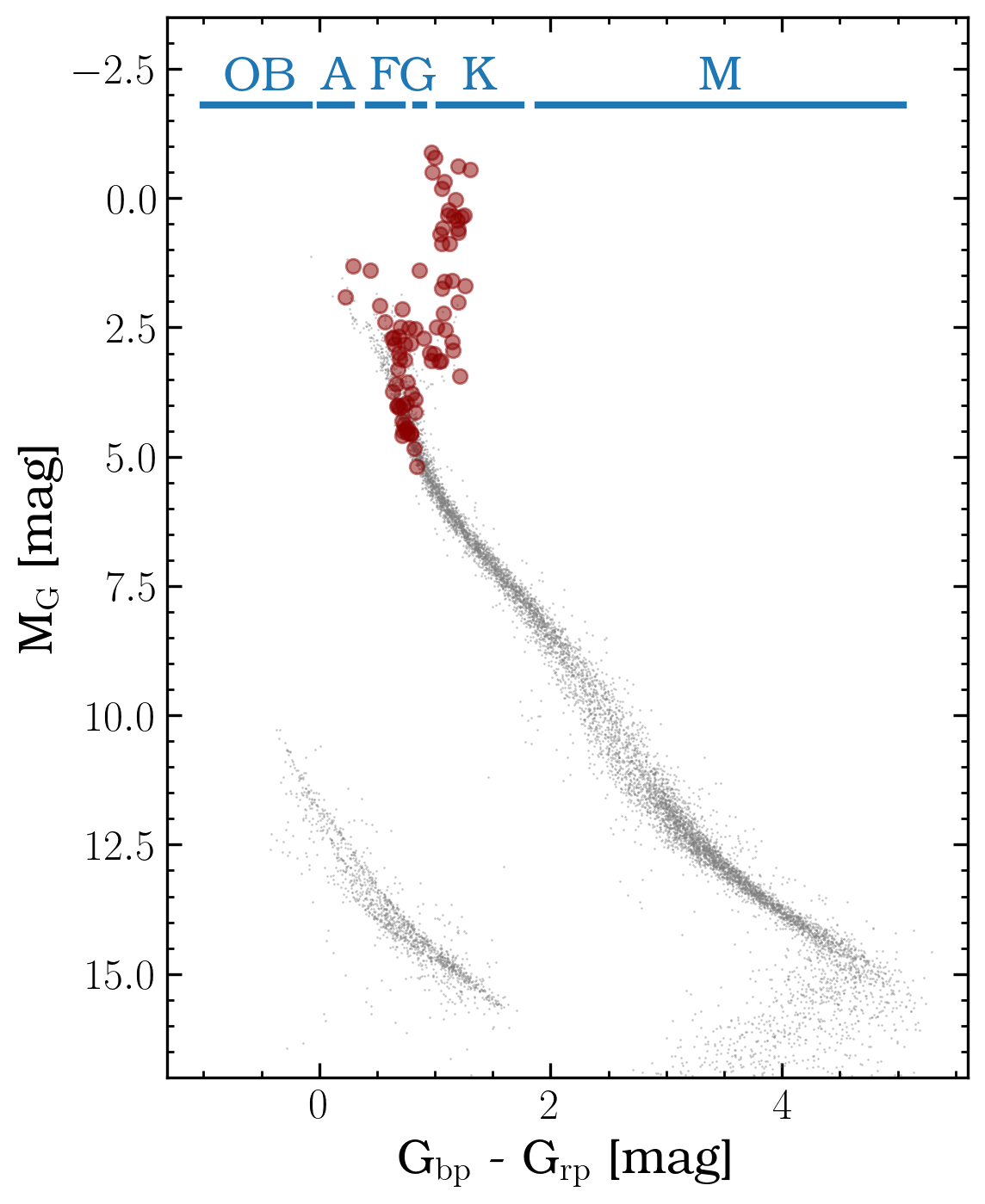}
\caption{\small{Top: Targets in RA/Dec. Targets accessible to MagAO-X from Las Campanas Observatory are plotted in blue.  Targets outlined by the purple stars were observed in the second observing semester (B) of 2022, and the yellow stars in the first semester (A) of 2024, and magenta stars in second semester 2024.  Bottom: Pup Search targets color magnitude diagram in \textsl{Gaia} absolute G magnitude vs \textsl{Gaia} bp - rp color.  The red points mark the 84 wide candidate WDMS systems identified in Sec \ref{sec: targets}.  The grey points are the 10,000 nearest high-quality solutions (RUWE $<$ 1.1) in \textsl{Gaia} DR3, with approximate spectral type ranges marked in blue (adapted from the spectral type-color relations in \citealt{PecautMamjek2013MSStarColors} \footnote{\url{http://www.pas.rochester.edu/~emamajek/EEM_dwarf_UBVIJHK_colors_Teff.txt}}).  
All of our target MS stars fall within the AFGK region, and $\approx$60\% are in the giant star region.
}}
\label{fig:master target cmd}
\end{figure}

Figure \ref{fig:master target cmd} (bottom) shows a \textsl{Gaia} color-magnitude diagram of our 84 target MS star candidate WD companion hosts.  All of our targets fall within the AFGK region of the CMD, and approximately 60\% fall in the giant branch region. To prioritize targets we use the star's \textsl{Gaia} renormilized unit weight error (RUWE, \citealt{lindegren_re-normalising_2018}) as a guide for multiplicity. \textsl{Gaia} RUWE encapsulates in a single number all deviation from \textsl{Gaia}'s assumed single star model, and RUWE $\gtrsim$1.2 could indicate multiplicity within the  range of \textsl{Gaia}'s sensitivity \citep{belokurov_unresolved_2020, Andrew2022RUWE, ElBadry2024GaiaBinaries}. \added{RUWE~$\approx$~1.0 could indicate that the source of the UV signal is widely separated, has a period shorter than the observational baseline \citep[1034 days, ][]{GaiaDR3-2023}, or the components are of nearly equal luminosity \citep{Penoyre2020BinaryDeviations}. As an additional metric, we employed estimates of proper motion anomaly between \textsl{Hipparcos} and \textsl{Gaia} from \citet[][]{Brandt2021HGCAEDR3} and \citet{Kervella2019}, which provides an estimate of the mass of a perturbing companion as a function of separation; approximately 30\% of our target list was observed with both missions. MagAO-X has demonstrated sensitivity to companions as close as $\approx$100--200~mas \citep{Males2022MagAOX, Males2024GMagAOX} in visible wavelengths, so we expect our survey to have sensitivity on the edge of the \textsl{Gaia} resolution and RUWE ranges, and we prioritized targets with RUWE~$\approx$~1.0--1.5}

\added{For a subset of our targets, \citetalias{Ren2020AFGKSample} estimated the WD and main sequence star \Teff\ from SED fitting and high resolution spectroscopy. We report those values where available.
}

\subsection{Observations}\label{sec: observations}
Targets observed for this work are summarized in Table \ref{tab:stellarparams}. We observed 3 Pup Search targets in 2022 and 18 in 2024 with the ExAO instrument MagAO-X \citep{Males2022MagAOX} on the 6.5m Magellan Clay Telescope at Las Campanas Observatory in Chile. MagAO-X is equipped with two science cameras, enabling simultaneous observation in two filter bands. The science cameras employ electron-multiplying CCDs (EMCCDs) to increase the number of electrons generated from a photon which reduces the effective readnoise. All camera settings were chosen to maximize signal while not saturating the target star. Observation details are summarized in Appendix A in Table \ref{tab:observations}

We observed in four broadband science filters: $g^\prime$ ($\lambda_0 = 0.527 \mu$m, $\Delta \lambda_{\rm{eff}} = 0.044 \mu$m), $r^\prime$ ($\lambda_0 = 0.614 \mu$m, $\Delta \lambda_{\rm{eff}} = 0.109 \mu$m), $i^\prime$ ($\lambda_0 = 0.762 \mu$m, $\Delta \lambda_{\rm{eff}} = 0.126 \mu$m), and $z^\prime$ ($\lambda_0 = 0.908 \mu$m, $\Delta \lambda_{\rm{eff}} = 0.130 \mu$m)\footnote{ Filter specifications and filter curves can be found in the MagAO-X instrument handbook at \url{https://magao-x.org/docs/handbook/index.html}}. We did not use $g^\prime$ images for any system in our analysis due to insufficient sensitivity. The pixel scale is 5.9 mas pixel$^{-1}$ \citep{Long2025AstrCalib}, and all science and dark frames were 1024$\times$1024 pixels (6\arcsec$\times$6\arcsec).

\textit{2022 December}. Conditions in 2022 were variable and we were only able to obtain usable images for 3 of the 2022 targets (TYC~4831-473-1, TYC~169-1942-1, and TYC~1262-1500-1).  All targets were observed in $i^\prime$ and $r^\prime$ simultaneously with the settings as tabulated in Table \ref{tab:observations}. 

\textit{2024 March -- 2024 May}. We observed 13 Pup Search targets in 2024 March and one in 2024 May.  All observations were obtained in $z^\prime$ and $i^\prime$ simultaneously, and $r^\prime$ and $g^\prime$ simultaneously with settings as tabulated in Table \ref{tab:observations}.
Seeing was variable across observations and ranged from 0.4--0.9\arcsec and conditions were excellent for most observations in 2024 March. Seeing and cloud cover hindered the observations in 2024 May.

\textit{2024 November}
We observed four targets in 2024 November. All targets were observed in $z^\prime$ and $i^\prime$ simultaneously. Seeing ranged from 0.7--1.2\arcsec, which limited observing to only the brightest accessible targets for AO correction.

\subsection{Data Reduction}\label{sec:datareduction}

We made use of multiple reduction methods for removing the stellar point-spread function (PSF) depending on the observation.  Which method was applied to each system is tabulated in Table \ref{tab:stellarparams}. Here we describe how each method was accomplished.

\textit{Radial Profile Subtraction} (RPS): For some systems, the fainter companion was bright enough and well-separated enough to be readily visible on-sky before any PSF subtraction.  For these observations, we frame-selected 700-800 of the best images, dark subtracted, registered, derotated, and summed the images.  We estimated the background as the median of an annulus around the central star outside the companion and deformable mirror speckles using \texttt{photutils}, \citep{Photutils_bradley_2023_7946442} and subtracted the background.
We then subtracted the host star's radial profile to remove stellar halo at the location of the companion.

\textit{Karhunen-Lo\`eve Image Processing + Angular Differential Imaging} (KLIP ADI): Systems for which a companion was not readily visible in imaging, and for which we had adequate sky rotation, were reduced using the principal component analysis (PCA) based Karhunen-Lo\`eve Image Processing (KLIP; \citealt{Soummer2012KLIP, AmaraQuanz2012PynPoint}) method with angular differential imaging (ADI, \citealt{Marois2006ADI}) using the python package \texttt{pyklip} \citep{Wang2015pyKLIP}.  Briefly, ADI observations involve allowing the field to rotate during observations, such that astrophysical sources rotate from image to image while optical artifacts (speckles, diffraction spikes, deformable mirror speckles) remain fixed. For each image in a dataset, KLIP produces an eigenimage basis set from the other images, then projects the target image onto the first $K$ eigenimages of the basis set (where $K$ is an integer) to create an estimated PSF. Assuming adequate sky rotation companion signals should not be in the estimated PSF. The estimated PSF is then subtracted from each image, the image is rotated to north-up-east-left, and the resulting image cube is combined to produce the final image.  
For each KLIP reduced dataset we selected $\approx$700 frames that span the observing time but allow rotation between frames to mitigate self-subtraction.

\textit{Classical Angular Differential Imaging} (Classical ADI):  We also made use of classical ADI \citep{Marois2006ADI} to search for candidate signals, particularly at large separations where KLIP is computationally inefficient. We performed classical ADI by selecting the best frames spanning the observation time, and produced a PSF estimate from the median of those frames. We subtracted the estimated PSF from each frame, then rotated and combined the frames into a single image. We also applied an unsharp mask to bring out any candidate signals. 

\textit{Reference Differential Imaging} (RDI): For some targets we were unable to obtain much sky rotation during the observation due to observing constraints. We reduced these datasets using reference differential imaging (RDI), in which we produced a reference PSF from another Pup Search target observed the same night in similar seeing conditions. For these datasets we made an estimated PSF as the median image of the reference star dataset, then for each image in the target star dataset we scaled the estimate to match the image, subtracted the estimated PSF from the image, rotated and combined the dataset, and applied an unsharp mask to the final image.

\floattable
\begin{longrotatetable}
\begin{deluxetable*}{cccccccccccc}
\tablecaption{{Stellar Parameters, Data Reduction, and Detections} \label{tab:stellarparams}}
\tablewidth{0pt}
\tablehead{
\colhead{TYC} & \colhead{Simbad Name} & \colhead{Obs Date}  & \colhead{R.A.} & \colhead{Decl.} & \colhead{SpT\tablenotemark{a}} & \colhead{MS \Teff\tablenotemark{b}} & \colhead{WD \Teff\tablenotemark{b}} & \colhead{RUWE} & \colhead{Distance\tablenotemark{c}} & \colhead{Red. Type\tablenotemark{d}} & \colhead{Comp.} \\
\colhead{} & \colhead{}& \colhead{} &
\colhead{} & \colhead{} & \colhead{} & \colhead{[K]} & \colhead{[K]} & \colhead{} & \colhead{[pc]} & \colhead{} & \colhead{Detected} 
}
\startdata
4831-473-1 & -- & 2022-12-14 & 07\h44\m37\s70 & -00\degree02\arcmin13\arcsec.87 & $^{*}$ & 5870$^{+24}_{-22}$ & 29740$\pm$2500 & 1.14 & 121.2$^{+0.2}_{-0.3}$ & RPS &  Yes \\
  & & 2024-03-22 &  & & & & & & & RPS &  Yes \\
169-1942-1 & -- & 2022-12-14 & 07\h25\m19\s38 & +03\degree31\arcmin03\arcsec.18 & $^{*}$ & 5050$^{+20}_{-16}$ & 29640$^{+2500}_{-2300}$ & 2.06 & 415$\pm$7 &  KLIP ADI & No \\
 & & 2024-03-26 &  & & & & & & & KLIP ADI &  No \\
1262-1500-1 & -- & 2022-12-14 & 04\h02\m13\s85 & +20\degree54\arcmin19\arcsec.98 & $^{*}$ &  -- & -- & 1.37 & 453$\pm$7  & KLIP ADI  & No \\
5480-589-1 & HD 87147 & 2024-03-21 & 10\h02\m54\s89 & -11\degree13\arcmin33\arcsec.14 & $^{*}$ & 4840$\pm$50 & -- & 1.42 & 632$\pm$13 & KLIP ADI &  No \\
4913-1224-1 & HD 92588 & 2024-03-22 & 10\h41\m24\s18 & -01\degree44\arcmin29\arcsec.37 & G9IV$^{1}$ & -- & -- & 1.11 &  37.66$\pm$0.04 & RPS &  Yes \\
1385-562-1  & BD+20 2007 & 2024-03-22 &  08\h08\m40\s15 & +20\degree17\arcmin02\arcsec.97 & $^{*}$ & 5856$\pm$3 & 25000$^{+1200}_{-1400}$& 0.82 & 118.2$\pm$0.3 & RPS &  Yes \\
4865-655-1 & BD-00 2082 & 2024-03-22 &  08\h52\m39\s93 & -01\degree00\arcmin47\arcsec.37 & $^{*}$& 4762$^{+10}_{-9}$& 40770$^{+3700}_{-3600}$ & 0.84 & 202.7$\pm$0.6 & RPS &  Yes \\
1451-111-1 & -- & 2024-03-22 & 13\h15\m14\s66 & +17\degree28\arcmin53\arcsec.24 & $^{*}$ & 6300$^{+20}_{-16}$& 15500$^{+1000}_{-900}$ & 1.14 & 222$\pm$1 & RPS &  Yes \\
5512-916-1 & BD-09 3292 & 2024-03-22 &  11\h28\m11\s65 & -10\degree14\arcmin21\arcsec.11 & $^{*}$ & -- & -- & 3.04 & 415$\pm$11 & KLIP ADI &  No \\
6712-1511-1 & CD-28 10038 & 2024-03-22 & 13\h22\m33\s50 & -28\degree40\arcmin07\arcsec.80 & $^{*}$ & -- & -- & 1.09 & 174.1$\pm$0.6& RDI &  No \\
877-681-1 & HD 109439 & 2024-03-23 & 12\h34\m37\s786 & +11\degree18\arcmin03\arcsec.51 & G5$^{2}$ & 5045$^{+21}_{-18}$ & 11740$^{+150}_{-140}$ &  1.00 & 260$\pm$2& RDI &  No \\
5518-135-1 & HD 104018 & 2024-03-23 & 11\h58\m41\s45 & -08\degree00\arcmin28\arcsec.38 &  G6/8IV$^{1}$& -- & -- & 1.14 & 122.3$\pm$0.5 & KLIP ADI &  Yes \\
288-976-1 & HD 108738 & 2024-03-27 &  12\h29\m33\s69 & +06\degree14\arcmin25\arcsec.29 & G0$^{2}$ &  -- & -- &1.58 & 158.8$\pm$0.8 & RPS &  Yes \\
1447-1616-1 & -- & 2024-03-28 & 12\h18\m00\s47 & +21\degree14\arcmin49\arcsec.66 & F8$^{3}$ &  -- & -- &1.03 & 392$\pm$2 & KLIP ADI &  No \\
368-1591-1 & HD 146740 & 2024-05-18 & 16\h17\m43\s36 & +01\degree29\arcmin48\arcsec.16 & K0III$^{1}$ & 4705$\pm$8 & 24500$^{+2000}_{-1900}$ & 1.51 & 220$\pm$2 & KLIP ADI &  No \\
26-39-1 & HD 7918 & 2024-11-09 & 01\h18\m57\s43 & +07\degree25\arcmin49\arcsec.49 & G5$^{2}$ & 4694$\pm$9 & 15000$^{+1000}_{-900}$ & 1.15 & 295$\pm$3 & KLIP ADI & No \\
6423-1892-1 & CD-27 191 & 2024-11-09 & 00\h39\m14\s49 & -26\degree25\arcmin19\arcsec.64 & $^{*}$ & -- & -- & 1.00 & 251$\pm$1 & KLIP ADI & No \\
5-436-1 & HD 680 & 2024-11-12 & 00\h11\m05\s65 & +04:\degree07\arcmin59\arcsec.13 & G8/K0III$^{1}$ & 4811$^{+7}_{-8}$ & 49000$^{+4700}_{-4600}$ & 1.36 & 267$\pm$3 & KLIP ADI & No \\
4685-1113-1 & BD-01 220 & 2024-11-21 & 01\h37\m27\s76 & -01\degree02\arcmin15\arcsec.82 & $^{*}$ & 4763$^{+10}_{-9}$ & 41000$^{+3700}_{-3600}$ & 2.38 & 596$\pm$13 & KLIP ADI & No \\
\enddata
\tablenotetext{a}{$^{*}$Spectral type not listed in Simbad, but its position on the CMD shown in Figure \ref{fig:master target cmd} places it in the red giant branch (RGB). Spectral Type references: $^{1}$\citealt{Houk1999SpectralTypes}, $^{2}$\citealt{Cannon1993HDcat}, $^{3}$\citealt{Kharchenko2001AllSky}}
\tablenotetext{b}{As reported in \citetalias{Ren2020AFGKSample} Table A2 or A3 from high-resolution spectroscopy.}
\tablenotetext{c}{\citealt{BailerJones2021Distances}}
\tablenotetext{d}{Reduction Type. `RPS': host PSF removed via radial profile subtraction;\deleted{`cADI': "Classical" ADI, host PSF removed via median image;} `KLIP ADI': host PSF removed via Karhunen-Lo\`eve image processing \citep{Soummer2012KLIP}. See text for details.}
\end{deluxetable*}
\end{longrotatetable}

\section{Results}\label{sec:results}

\begin{figure*}
\centering
\includegraphics[width=0.45\textwidth]{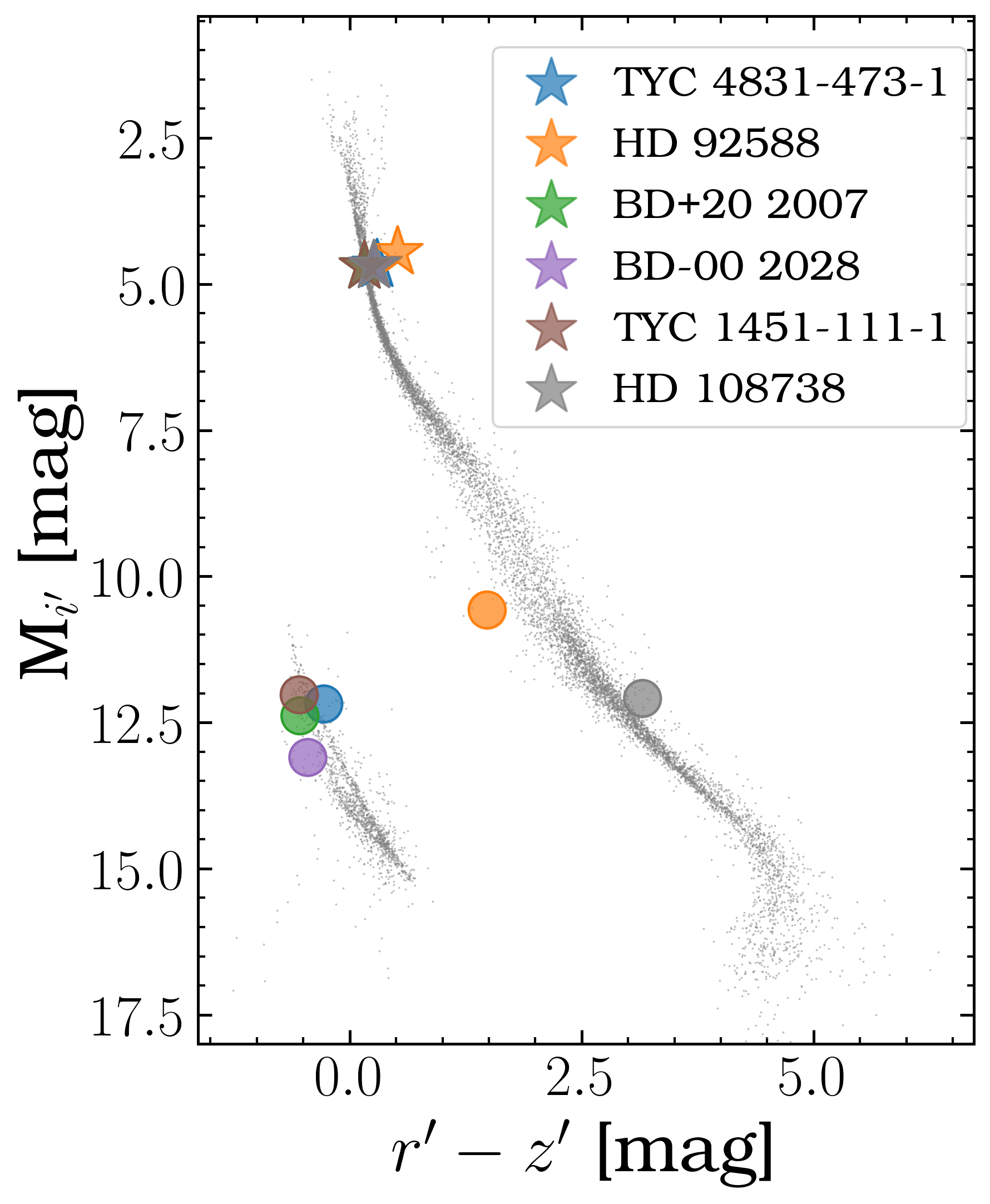}
\includegraphics[width=0.45\textwidth]{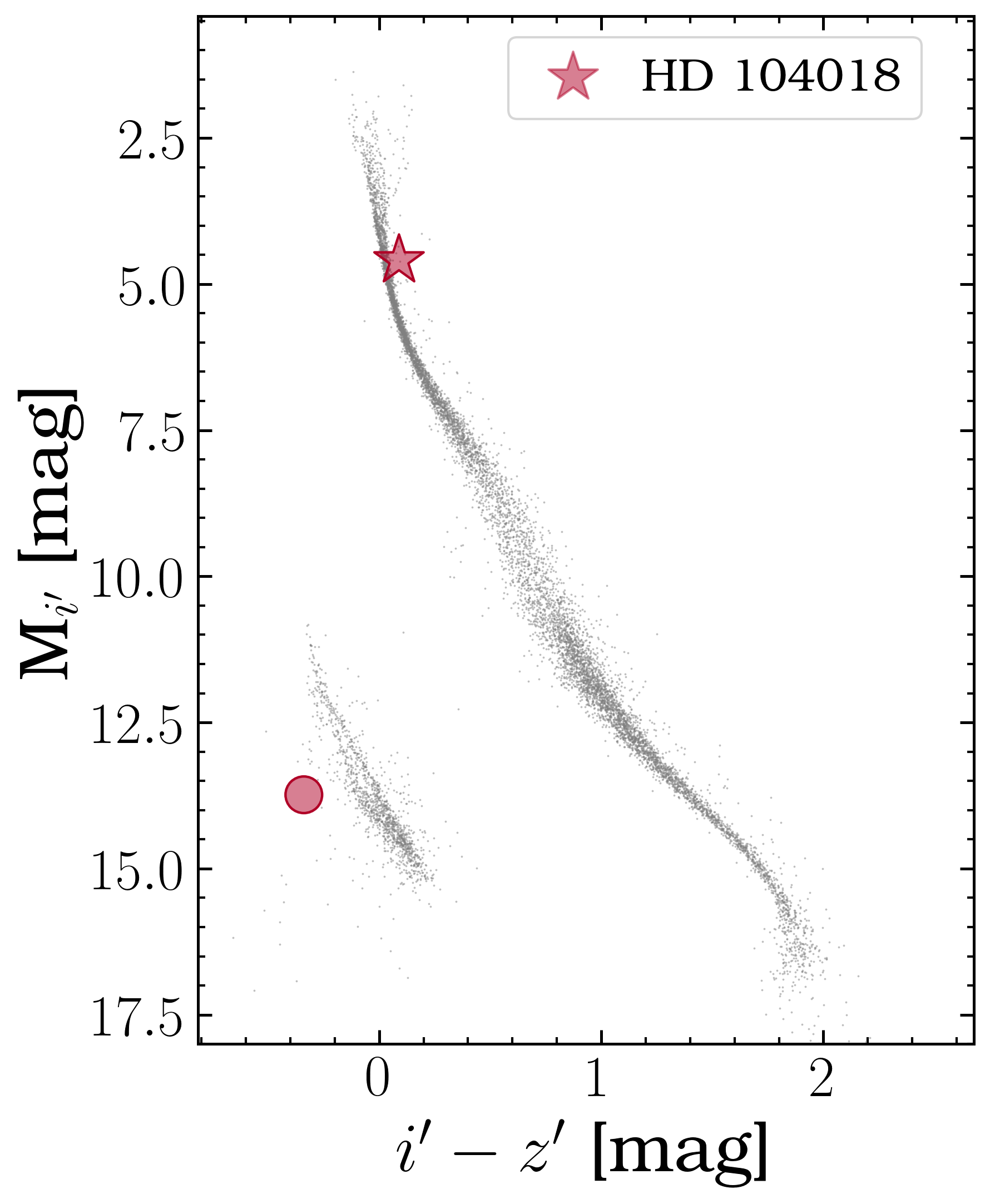}
\caption{\small{Left: Color-magnitude diagram of the confirmed and candidate companions from Table \ref{tab:comp-properties} shown in SDSS $r^\prime - z^\prime$ color and SDSS $i^\prime$ absolute magnitude. Colored stars give the model photometry of the host star, and colored circles show the measured photometry of the corresponding companion in MagAO-X filters converted to SDSS filters using color conversion table\footnote{\url{https://magao-x.org/docs/handbook/observers/filters.html}}. Grey dots are the same \textsl{Gaia} sample from Figure \ref{fig:master target cmd} converted to SDSS colors using \texttt{GaiaXPy} \citep{daniela_ruz_mieres_2023_GaiaXPy} to generate synthetic photometry from \textsl{Gaia} DR3 spectra.  Right: Same as left but $i^\prime - z^\prime$ color and SDSS $i^\prime$ absolute magnitude for the candidate companion signal to HD 104018, since we do not have $r^{\prime}$ photometry for that target. The candidate companion clearly falls in the WD regime, however the exact position in the WD sequence is not well constrained with only $i^{\prime}$ and $z^{\prime}$ colors.}
}
\label{fig:comp-cmd}
\end{figure*}

We detected six new candidate companions, identified in Table \ref{tab:stellarparams}. Five are identified as candidate companions from a single epoch of imaging. We have additionally identified a companion to TYC~4831-473-1 with two epochs and were able to establish companion status through common proper motion analysis (see Sec \ref{sec:tyc4831-473-1}). Companion properties are given in Table \ref{tab:comp-properties} and in the text that follows.

\subsection{Detections}\label{sec: detections}

For the seven candidate companion signals we measured their photometry, fit model spectra to our photometry, and measured the relative astrometry. In this section we describe our measurements and results. 

\subsubsection{Astrometry}

\textit{RPS reduced datasets:} To estimate companion separation and position angle, we performed a bootstrap simulation: 
\begin{enumerate}
    \item We selected a random subset of $N$ images, where $N < N_{\rm{total\; images}}$.
    \item We reduced the subset of $N$ images following the radial profile reduction described in Sec \ref{sec: targets}.
    \item We fit a 2D Gaussian model to the host star and companion signal and computed relative separation in mas and position angle east of north.
    \item We repeated this procedure $N_{tot}$ times and took the final separation and position angle as the mean and standard deviation from the $N_{tot}$ trials.
\end{enumerate}
We used $N=500$ and $N_{tot}=100$ for each of the candidate and confirmed companions.

\textit{KLIP reduced datasets:} For the signals discovered in KLIP reductions, we took the separation and P.A. of the negative injected signal described below as the mean value in each filter. To estimate the uncertainty we took the range of separation and PAs tested within one standard deviation of the minimum negative signal. We then averaged the mean in each filter and added the uncertainties in quadrature to attain the values reported in Table \ref{tab:comp-properties}

\subsubsection{Photometry}

\textit{RPS reduced datasets:} For each wide companion reduced via radial profile subtraction, we used the host star as the photometric reference star to obtain absolute photometry for the companion. The host star was not saturated in any dataset. To measure the photometry, we placed an aperture of radius 1$\lambda/D$ centered at the subpixel location of the host star in the background subtracted image and the the location of the companion signal in the radial profile subtracted image, and measured the signal within the apertures using the \texttt{photutils} aperture photometry package as the sum of pixel values in the aperture. We estimated the noise in the photometric apertures as the signal Poisson noise in the aperture added in quadrature to the background noise, estimated as the standard deviation of pixel values in an annulus around the central star beyond the stellar halo and deformable mirror speckles (radius 300 -- 350 pixels) multiplied by the aperture area. We then computed the flux contrast as the ratio of the companion to host photometric signal. We estimated the host star flux in each filter using stellar models from the Pickles atlas \citep{Pickles1998} corresponding to the spectral type reported in Table \ref{tab:stellarparams}, then applied our photometric contrast to estimate the companion flux. 

\textit{KLIP reduced datasets:} For the candidate signal discovered in the KLIP-reduced dataset (HD~104018), we estimated the photometry using negative signal injection. Into each image in the dataset we injected a negative PSF (a scaled and inverted median PSF from the dataset) at a known separation, position angle (P.A.), and contrast, reduced the dataset using the same KLIP configuration as the initial reduction, then measured the standard deviation in an aperture of diameter 1$\lambda/D$ at the candidate signal location. We repeated this for an array of separations, position angles, and contrasts and determined the injected signal which minimized the standard deviation in the reduced image. We took the known contrast of this injected signal as the companion contrast and estimated the companion flux in the same manner as above. We measured the signal to noise ratio using the method of \citet{Mawet2014} for estimating signal to noise ratio in the regime of small number of photometric apertures at close separations.  At the separation of the signal, there are N = 2$\pi r$ resolution elements of size 1~$\lambda/D$ (the characteristic scale of speckle noise), where $r = n \lambda$/D and n varies with the filter wavelength. We defined a ring of N-3 resolution elements (neglecting those at and immediately to each side of the signal) at separation $r$ with radius 0.5~$\lambda/D$, then applied Eqn (9) of \cite{Mawet2014}, which is the Student's two-sample t-test:
\begin{equation}\label{eq:snr}
    p(x,n2) = \frac{\bar x_1 - \bar x_2}{s_2 \sqrt{1 + \frac{1}{n_2}}}
\end{equation}
where $\bar x_1 = $ flux, $\bar x_2 =$ mean[$\Sigma$(pixels in apertures)], $s_2 = $ stdev[$\Sigma$(pixels in apertures)], n$_2$ = N-3, and S/N = p.  The denominator of that equation is the noise term.

\textit{Photometry results:} Candidate signal photometry is shown on a color-magnitude diagram in Figure \ref{fig:comp-cmd}. For companions in the white dwarf regime of the CMD, we fit our photometry with Montreal WD Model \citep{Tremblay2011ImprovedSpecAnalyOfDA-WD, Bergeron2011CompSpecAnalyOfDB-WD, blouinNewGenerationCool2018-I, Bedard2020WDModels} synthetic photometry for hydrogen and helium-dominated white dwarfs\footnote{Accessed from \url{https://www.astro.umontreal.ca/~bergeron/CoolingModels/} on July 30th 2024}. The synthetic photometry absolute magnitudes are provided for multiple photometric systems in (\Teff, $\log g$) bins, with \Teff\ from 1500--5500~K with 250~K spacing, 6000--17,000~K with 500~K spacing, 20,000--90,000~K with 5000~K spacing, and 100,000--150,000~K with 10,000~K spacing, and $\log g$ from 7.0--9.0 in bins of 0.5. 
For each WD candidate companion we used the following procedure:
\begin{enumerate}
    \item We generated a bootstrapped array of 1000 simulated observations drawn from our companion fluxes and uncertainty in each filter $z^\prime$, $i^\prime$, $r^\prime$ (we excluded $g^\prime$ due to poor photometry in that filter), converted to AB magnitudes, then applied color correction to convert from MagAO-X filters to SDSS filters to obtain 1000 SDSS apparent magnitudes and uncertainties from our observations.
    \item To account for distance uncertainty, we calculated model and observed WD absolute magnitudes at 20 parallaxes from a normal distribution described by the \textsl{Gaia} parallax measurements for the primary star. Each model (T$_{\rm{eff}}$, $\log g$) pair has 20,000 simulated observations to compare.
    \item We computed goodness-of-fit using a $\chi^2$ metric:
    \begin{equation}
        \chi^2 = \sum_f{\left( \frac{M_{d,f} - M_{m,f}}{\sigma_{f}} \right)}
    \end{equation}
where $f$ are the filters $z^\prime$, $i^\prime$, $r^\prime$; $M_{d,f}$ is the observation absolute magnitude in each filter; $M_{m,f}$ is the model absolute magnitude; and $\sigma_f$ is the uncertainty in each filter, taken from the signal-to-noise ratio (S/N) in each filter as $\sigma_f = \frac{F_{d,f}}{\rm{S/N}_f}$ and converted to magnitudes.
    \item we computed a total $\chi^2_{\rm{tot}} = \sum_f{\chi^2_f}$ for each simulated observation for that model.
    \item The final $\chi^2$ for each each model (T$_{\rm{eff}}$, $\log g$) pair is the mean and standard deviation of the $\chi^2_{\rm{tot}}$ of the simulated observations.
\end{enumerate}

By computing model fits to bootstrapped simulated observations spanning our uncertainties and \textsl{Gaia} parallax, we incorporated the uncertainties in both our observations and parallax, and obtained uncertainties on the $\chi^2$ value for each fit. This procedure assumes the companion is bound to the primary, which we have only established for TYC~4831-473-1~B to date (Section 3.1.3). Montreal model fits to the photometry of the new WDs is shown in Figure \ref{fig:WD-modelfits}.

For companions in the main sequence regime, we fit \texttt{Phoenix} stellar spectra \citep{Allard2012Phoenix3} to our photometry using \texttt{pysynphot} \citep{pysynphot-citation} and the following procedure: 

\begin{enumerate}
    \item We generated a bootstrapped array of 10000 simulated observations drawn from our photometry and uncertainty in each filter $z^\prime$, $i^\prime$, $r^\prime$ (we excluded $g^\prime$ due to poor photometry in that filter).
    \item We selected a filter and generated synthetic photometry for the model in that filter and scaled the model such that model photometry in that filter matched observed photometry.
    \item We computed new synthetic photometry from the model for each filter and computed goodness-of-fit using a $\chi^2$ metric:
    \begin{equation}
        \chi^2 = \sum_f{\left( \frac{F_{d,f} - F_{m,f}}{\sigma_{f}} \right)}
    \end{equation}
where $f$ are the filters $z^\prime$, $i^\prime$, $r^\prime$; $F_{d,f}$ is the simulated observation flux in each filter; $F_{m,f}$ is the scaled model flux in each filter; and $\sigma_f$ is the uncertainty in each filer, taken from the signal-to-noise ratio (S/N) in each filter as $\sigma_f = \frac{F_{d,f}}{\rm{S/N}_f}$.
    \item We computed a total $\chi^2$ and mean and standard deviation in the same manner as above.
\end{enumerate}

In scaling this way we fit the shape of the model to our photometry and do not include uncertainties in distance and radius (which are unknown). \texttt{Phoenix} model fits to our photometry are shown in Figure \ref{fig:MS-modelfits}

\subsubsection{TYC 4831-473-1 B}\label{sec:tyc4831-473-1}
\begin{figure*}
\centering
\includegraphics[width=0.95\textwidth]{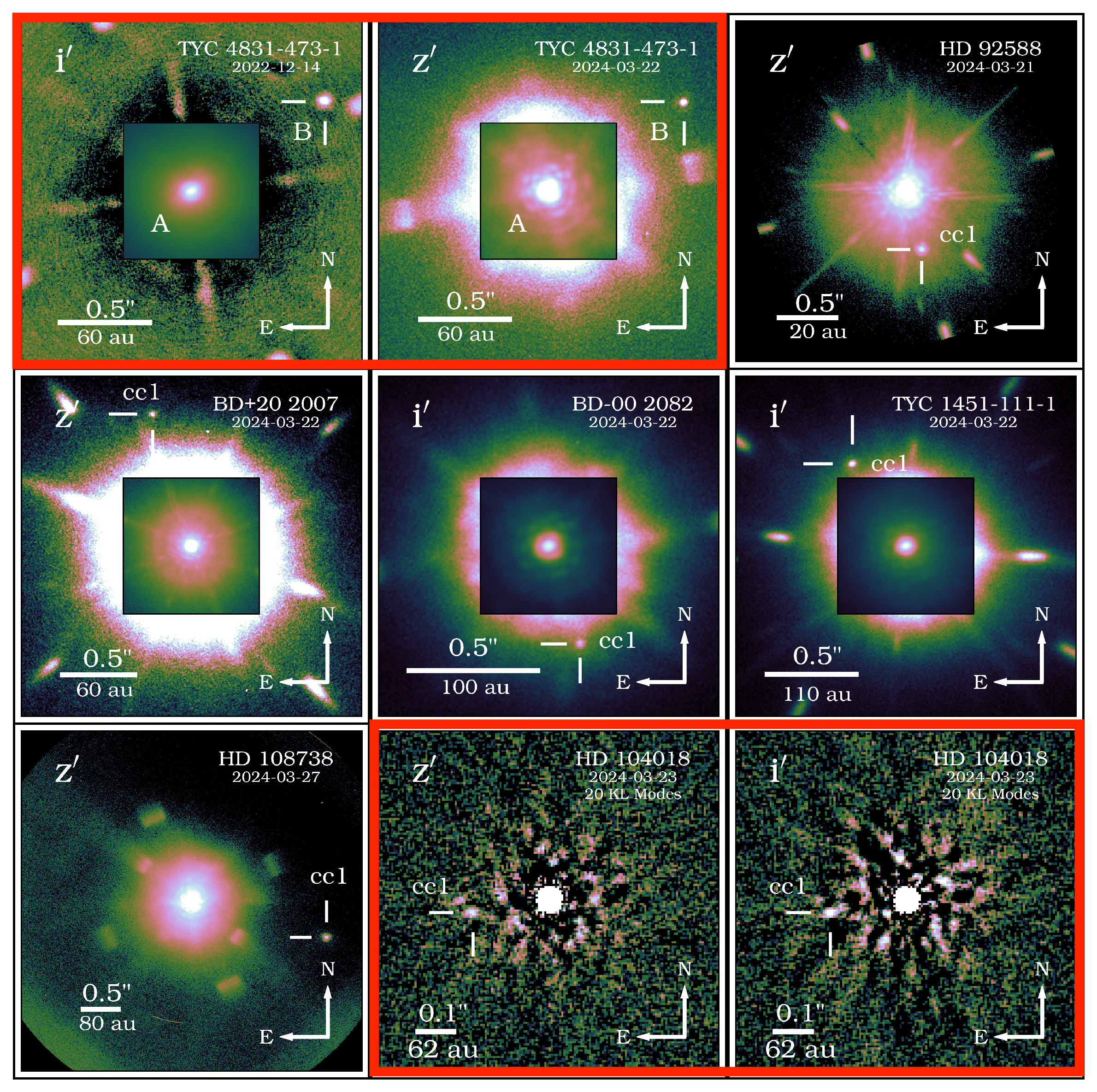}
\caption{\small{MagAO-X images of the confirmed and candidate companion signals in this work.  All images are North up/ East left; the filter, host star, and observation date are as indicated. Scale bars show angular and physical scales. The companion is marked by a white cross and label. TYC~4831-473-1~B displays two images (in the top red rectangle), $i^{\prime}$ from 2022 and $z^{\prime}$ from 2024, to show common proper motion. We show two images for HD~104018 (in the bottom red rectangle) in $z^{\prime}$ and $i^{\prime}$ filters to illustrate that candidate signal does not move between filters as it would if it were a speckle. \added{For the faint candidate signals near TYC~4831-473-1 (2024), BD+20 2007, BD-00 2028, and TYC~1451-111-1, we show the image in two stretches to highlight the faint companion with an inset showing the star at a different stretch.}
}}
\label{fig:comp-images}
\end{figure*}

We detected a candidate companion signal to the northwest of TYC~4831-473-1 in 2022 December and again in 2024 March. Fig \ref{fig:comp-images} shows an $i^\prime$ image from 2022 and a $z^\prime$ image from the 2024 observation, with the companion marked by a white cross $\approx$900~mas to the northwest. TYC~4831-473-1 is estimated to have a spectral type of $\approx$G2V based on \textsl{Gaia} DR3 colors and \citet{PecautMamjek2013MSStarColors} reference colors, with \textsl{Gaia} DR3 RUWE~=~1.14 (source id: 3085864943402832000). The companion's photometry falls in the WD sequence, as shown in Figure \ref{fig:comp-cmd}.  

\begin{figure*}
\centering
\includegraphics[width=0.9\textwidth]{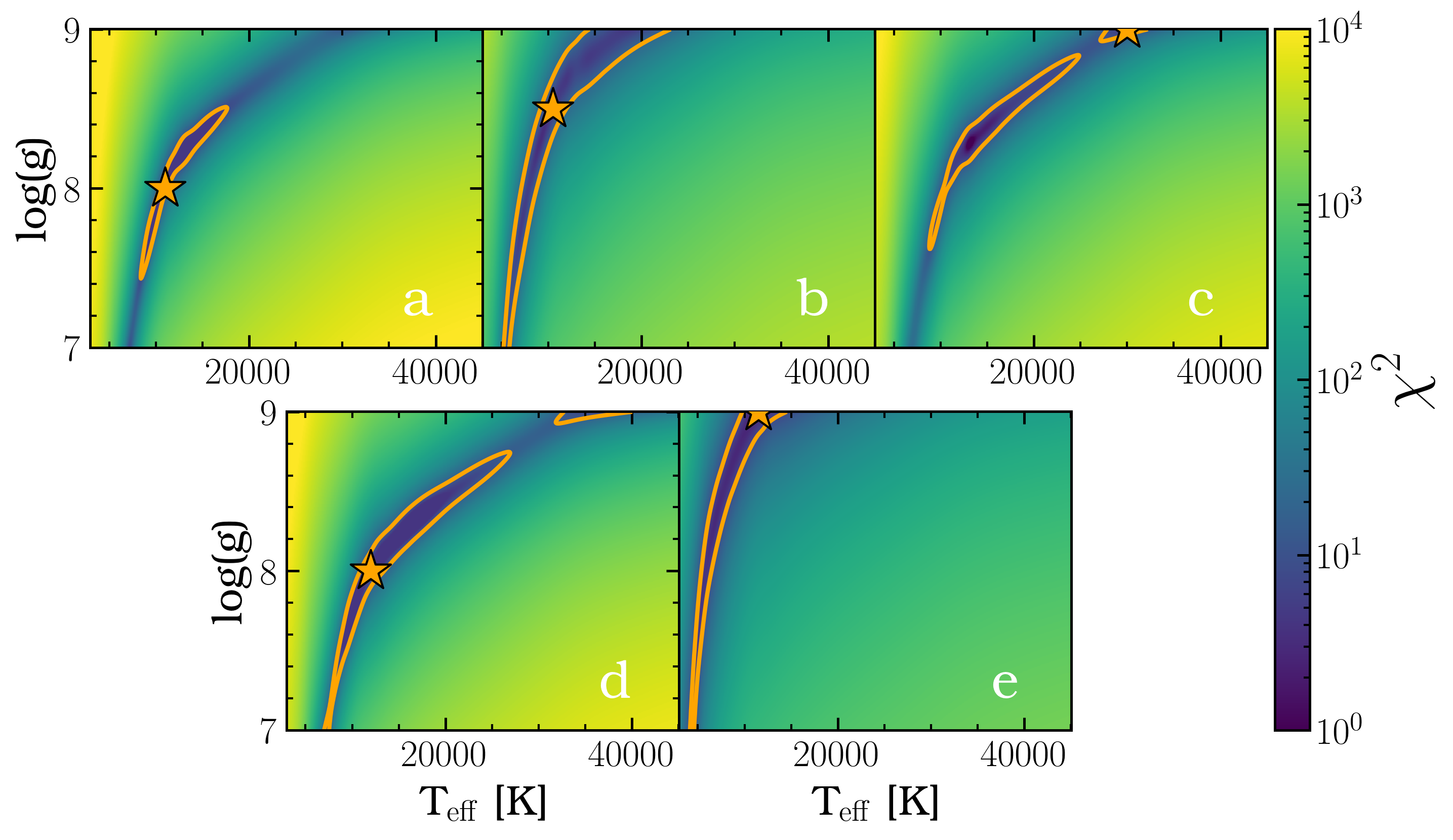}
\caption{\small{Map of $\chi^2$ as a function of \Teff\ and $\log g$ for hydrogen dominated WD atmosphere models fit to our photometry for the five candidate WD companions. The effective temperature is on the X-axis and $\log g$ on the Y-axis with $\chi^2$ value interpolated between grid points. The best fitting model is marked with an orange star; the solid contour shows the 2$\sigma$ surfaces from that minimum $\chi^2$ value. a: TYC~4831-473-1~B, b: BD+20~2007~cc1, c: BD-00~2082~cc1, d: TYC~1451-111-1~cc1, e: HD~104018~cc1
}}
\label{fig:WD-modelfits}
\end{figure*}

Our photometry rules out main sequence and helium-dominated WD atmospheres. Figure \ref{fig:WD-modelfits} (a) shows the $\chi^2$ surface for WD hydrogen-dominated atmosphere models to our photometry with values interpolated between model grid points. There is a relationship between \Teff\ and $\log g$ -- higher \Teff\ models fit better at higher $\log g$ and vice versa. The orange star marks the best fitting model grid point at \Teff~=~11000~K, $\log g$~=~8.0, with orange contours showing regions within 2$\sigma_{\chi^{2}}$ of that best fitting model. The lowest $\chi^2$ model has \Teff~=~11000~K, $\log g$~=~8.0, but in our final reported results we include all models within 2$\sigma_{\chi^{2}}$ of that model's $\chi^2$, and determine that TYC~4831-473-1~B has a H-dominated atmosphere with \Teff~=~8500~--~17000~K and $\log g$~=~7.5~--~8.5. \added{\citetalias{Ren2020AFGKSample} predicted the WD responsible for the UV excess to have \Teff~=~30000$\pm$2500~K from high resolution spectroscopy, which is not consistent with our results. Further spectroscopic characterization of the now-resolved companion is warranted.}

\subsubsection{TYC 4831-473-1 B Common proper motion}

With two epochs of relative astrometry spanning $\approx$1.2 years, we were able to confirm companion status.  Figure \ref{fig:PupS1B-cpm} shows the common proper motion of the companion relative to the host star.  The black track shows the expected path of the companion relative to TYC~4831-473-1 if it were a non-moving background object; the circles mark the observed offset from the host star in the two epochs. Relative to the 2022 epoch, if the companion were a non-moving background object we would expect to observe it at the location of the red diamond in the 2024 epoch; instead we observe it at the location of the red circle.  Our observed motion is not consistent with a background object, so we conclude that TYC~4831-473-1~B is not a background object and posit that the observed motion is due to orbital motion. Future observations are needed to constrain the orbit.

\begin{figure*}
\centering
\includegraphics[width=0.9\textwidth]{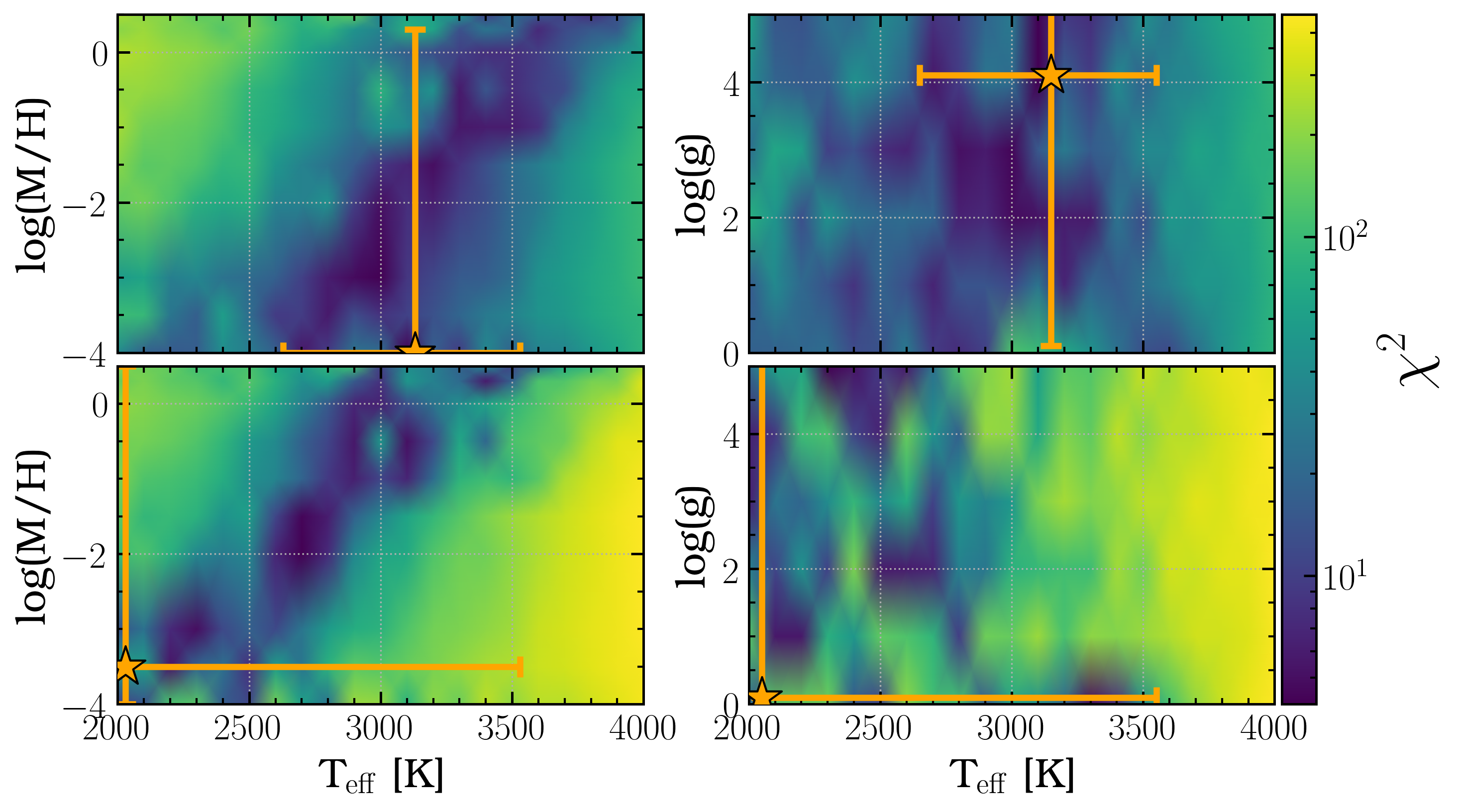}
\caption{\small{Map of $\chi^2$ as a function of \Teff, log([M/H]), and $\log g$ of Phoenix main sequence star models to our photometry for HD~92588~cc1 (top) and HD~108738~cc1.  The effective temperature is on the X-axis and log([M/H]) (left) and $\log g$ (right) on the Y-axis with $\chi^2$ value interpolated between grid points. The best fitting model is marked with an orange star; the error bars show the 1$\sigma$ values from that minimum $\chi^2$ model. Our photometry is unable to constrain log([M/H]) or $\log g$ for either signal.
}}
\label{fig:MS-modelfits}
\end{figure*}

\begin{figure}
\centering
\includegraphics[width=0.45\textwidth]{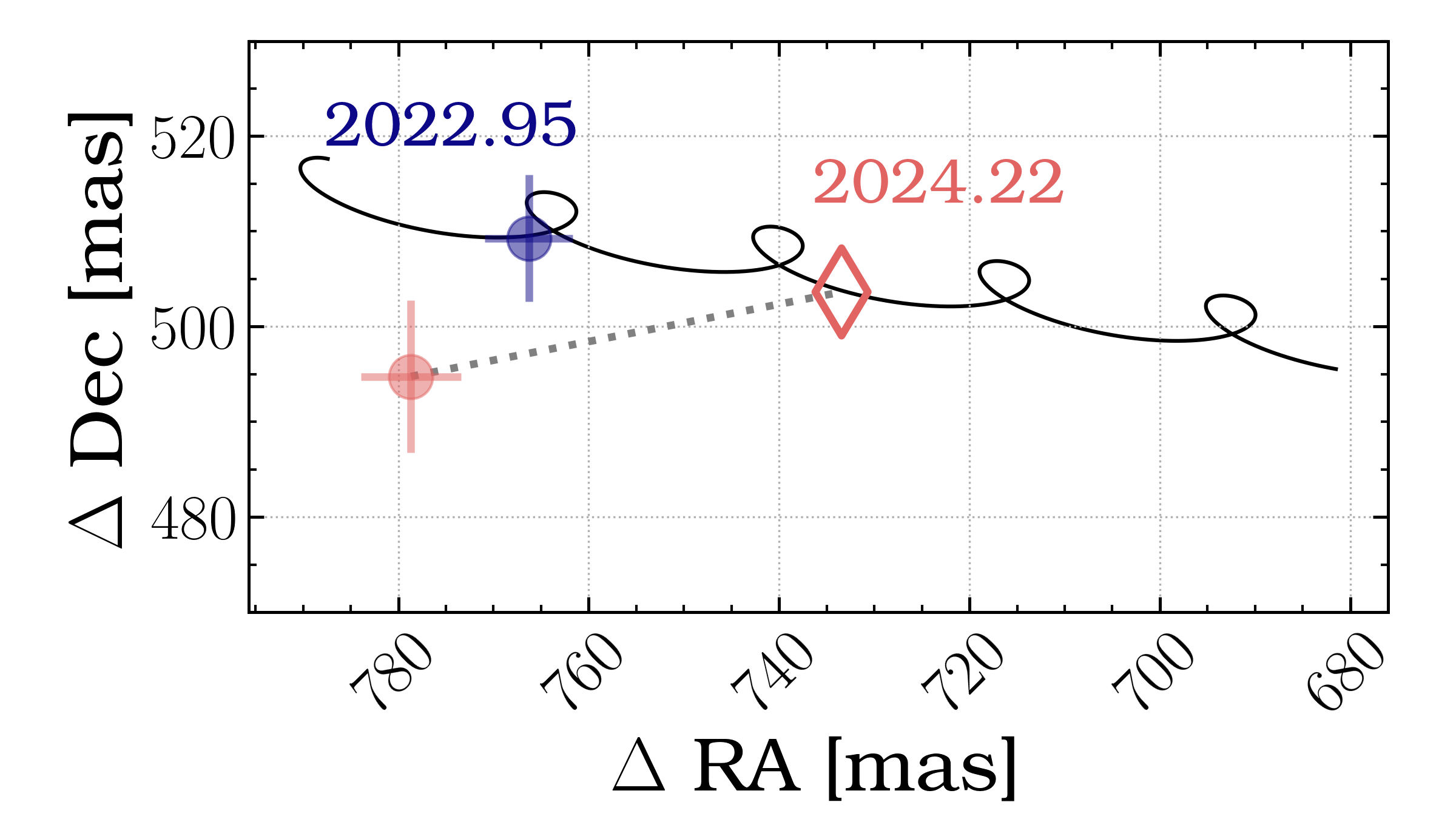}
\caption{\small{Common proper motion plot for TYC~4831-473-1~B. The offset from TYC 4831-473-1 in RA and Dec are given on the x- and y-axis respectively, with our observed position in 2022 and 2024 are given by the blue and red circles.  If TYC~4831-473-1~B were an unmoving background star, it would follow the black track relative to TYC~4831-473-1, given by the host star's proper motion and parallax; we would have observed TYC~4831-473-1~B at the location of the red diamond in 2024.  Our observed location is highly discrepant with this predicted location and moving in the opposite direction in RA.  We conclude that TYC~4831-473-1~B is not a background star, and posit that the apparent motion of TYC~4831-473-1~B is due to orbital motion.  More observations are required to begin to pin down orbital parameters.
}}
\label{fig:PupS1B-cpm}
\end{figure}

\subsubsection{HD~92588~cc1}

HD~92588~candidate companion number one (cc1) is a candidate signal $\approx$500~mas (20~au) to the southwest of the G9IV \citep{Torres2006SACYi}, subgiant \citep{Jofre2015StellarParameters} star HD~92588 (aka 33 Sex, TYC~4913-1224-1, \textsl{Gaia} source id: 3805647792295621376). It falls below the main sequence as shown in Figure \ref{fig:comp-cmd}, suggesting it is bluer than a main sequence star. We posit that HD~92588~cc1 is an M-dwarf + WD unresolved binary. If it is a gravitationally bound system, HD~92588~cc1 does not fit the properties of a cool subdwarf \citep{Gizis1997MSubdwarfs}. HD~92588 is estimated to be 5 Gyr old giant star \citep{Tsantakie2013StellarParameters, Jofre2015StellarParameters, Delgado2019HarpsAbundances, GaiaDR3-2023} with solar metallicity \citep{Jofre2015StellarParameters} and classified as a metal rich Galactic disk star \citep{CostaSilvia2020ChemicalAbuncances}. These characteristics are not consistent with the old age, subsolar metallicity, halo star subdwarf population (\citealt{Lepine2007SubdwarfClassifications}). Thus we hypothesize there is an unresolved WD companion to a cool main sequence star responsible for the abnormally blue color. Spectroscopic follow-up is needed to determine its nature, and astrometric time series are needed to establish relative motion between HD~92588 and HD~92588~cc1.\deleted{; HD~92588 is a high-proper motion star it shouldn't take too long to establish common proper motion. As HD~92588 is well characterized, astrometric time series should be able to constrain companion mass.}

Due to the blue color, \texttt{Phoenix} models did not fit our photometry well, with the lowest $\chi^2$ value being $\approx$18. \deleted{Our photometry was bluer than the Phoenix models, with the $r^\prime$ being consistently brighter than the models.} We estimate this star to be approximately a mid-M dwarf, but further careful photometry and spectroscopy is needed to determine its spectral type.  While it is not a new SLS system, it merits further followup as an interesting object.

\begin{figure}
\centering
\includegraphics[width=0.45\textwidth]{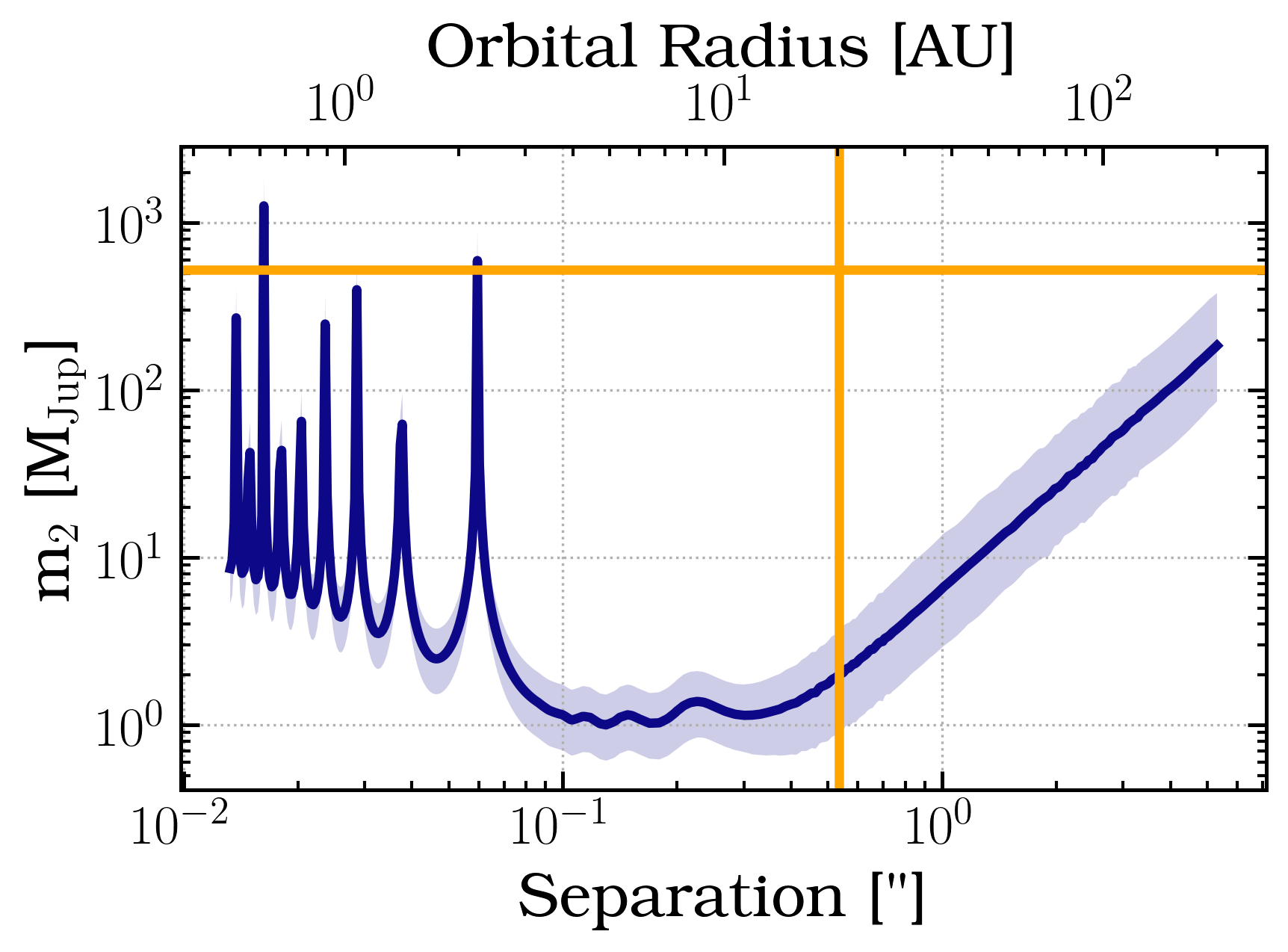}
\includegraphics[width=0.45\textwidth]{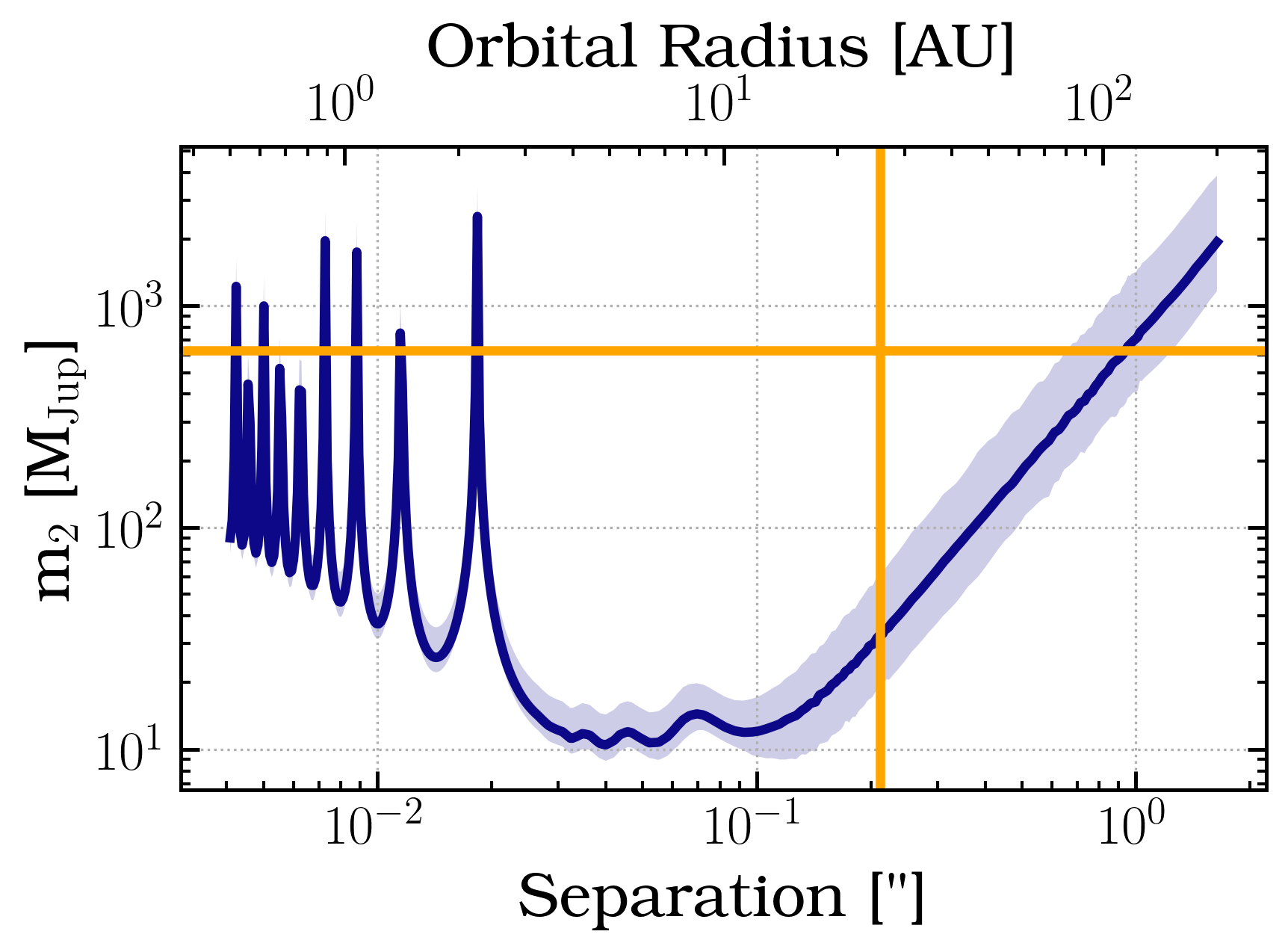}
\caption{\small{Proper motion anomaly (PMa) plot for HD~92588 and HD~104018 using the methodology of \cite{Kervella2022}. This plot describes the (minimum) mass of an object as a function of separation that would produce the observed proper motion anomaly between the \textsl{Hipparcos} and \textsl{Gaia} observations.  Top: The orange vertical line marks the separation of HD~92588~cc1 from HD~92588, and the horizontal orange line marks the approximate mass of a mid-M dwarf. If HD~92588~cc1 is an M dwarf-white dwarf binary, it would lie even higher on this plot.  Bottom: The orange vertical line marks the separation of HD~104018~cc1 from host, and the horizontal vertical line marks the approximate mass of a white dwarf.  Both companions fall well above the sensitivity curve, suggesting that neither is responsible for the observed acceleration and may be unassociated.
}}
\label{fig:PupScc-pma}
\end{figure}

Figure \ref{fig:PupScc-pma} (top) shows the proper motion anomaly (PMa) curve for HD~92588 (generated using the catalog and methodology of \citealt{Kervella2022}), indicating the mass sensitivity as a function of separation for the observed acceleration of the star between the \textsl{Hipparcos} and \textsl{Gaia} astrometric measurements; the orange vertical line marks the separation of HD~92588~cc1, and the horizontal line marks the approximate mass of a mid-M dwarf star.  Its position is well above the sensitivity curve. Additionally, in the \textsl{Hipparcos-Gaia} Catalog of Accelerations \citep[HGCA, ][]{Brandt2021HGCAEDR3} the PMa has a reduced $\chi_{\nu}^2 = 1$, indicating the motion is consistent with a single star. HD~92588~cc1 may not be a bound companion, and it will be included in future observations for a common proper motion analysis. Even if unassociated, it presents a unique opportunity to study a WD+dM close binary at high spatial resolution close to a bright natural AO guide star, without the need for laser guide star AO. \deleted{indicates that either cc2 is on a highly-inclined or highly eccentric orbit, or there is another unresolved object bound to 33 Sex contributing to the PMa and cc2 is actually a chance alignment and not bound.  Followup is warranted for this system to observe the orbit and search for closer-in companions.}

\subsubsection{BD+20 2007 cc1}

BD+20~2007~cc1 is a candidate companion signal $\approx$950 mas (110 au) to the north east of BD+20~2007 (TYC~1385-562-1, \textsl{Gaia} source id: 669892921105281024) in the white dwarf region of Figure \ref{fig:comp-cmd}. BD+20~2007  is estimated to have a spectral type of $\approx$G0 based on \textsl{Gaia} DR3 colors and \citet{PecautMamjek2013MSStarColors} reference colors; its \textsl{Gaia} DR3 RUWE is 0.83.  

Our photometry rules out helium-dominated WD and main sequence star models, and is best fit by H-dominated WD models. The $\chi^2$ surface and best fitting models are shown in Figure \ref{fig:WD-modelfits} (b). The best fitting model H-dominated WD model has \Teff~=~10000~K and $\log g$~=~8.5, but models from \Teff~=~6000~--~12000~K and $\log g$ from 7.0~--~9.0 have $\chi^2$ values within 2$\sigma$ of that best fit, so we report these values. \citetalias{Ren2020AFGKSample} predicted the WD responsible for the observed UV excess to have \Teff~=~25000$^{+1400}_{-1200}$~K, which is hotter than our results.

\subsubsection{BD-00 2028 cc1}

BD-00 2028 cc1 is a candidate companion signal $\approx$400 mas (80 au) south west of BD-00 2028 (TYC~4865-655-1, \textsl{Gaia} source id: 3073915309292882816), an estimated F8 star with RUWE~=~0.83. Our photometry is best fit by H-dominated WD models. The $\chi^2$ surface and best fitting models are shown in Figure \ref{fig:WD-modelfits} (c). The best fitting model H-dominated WD model has \Teff~=~30000~K and $\log g$~=~9.0, and models from (10500, 7.75) -- (30000, 9.0) are within 2$\sigma$ of that model.

\subsubsection{TYC~1451-111-1~cc1}

TYC~1451-111-1~cc1 is a candidate companion signal $\approx$550 mas (120 au) north east of TYC~1451-111-1 (\textsl{Gaia} source id: 3937175633819731968), an estimated F8 star with RUWE~=~1.14. Our photometry rules out helium-dominated WD and main sequence models.  The $\chi^2$ surface and best fitting models are shown in Figure \ref{fig:WD-modelfits} (d). The best fitting model occurred at (\Teff,$\log g$) = (12000~K, 8.0), with models from (7000~K, 7.0) to (40000~K, 9.0) within 2$\sigma_{\chi^2}$.\citetalias{Ren2020AFGKSample} predicted the WD responsible for the UV excess would have \Teff~=~15,500$^{+1000}_{-900}$~K from high resolution spectroscopy, which is consistent with our results.

\subsubsection{HD~108738~cc1}

HD~108738~cc1 is a candidate companion signal $\approx$2\arcsec\ (120 au) southwest of TYC~288-976-1 (HD 108738, \textsl{Gaia} source id: 3708538031977678976), a G0 star with RUWE~=~1.58. It falls in the late-M dwarf section of the CMD in Figure \ref{fig:comp-cmd}.  Figure \ref{fig:MS-modelfits} shows the results of our \texttt{Phoenix} model fits. Our photometry does not constrain metallicity or $\log g$ and constrains \Teff~$<$~3500~K, so we estimate it is a mid- to late M dwarf ($\lesssim$~M2).

Since HD~108738~cc1 is approximately a mid-M dwarf, it is not the source of the GALEX excess. There is no PMa for this system as it was not observed in \textsl{Hipparcos}. The elevated RUWE suggests another companion closer than the one we detected, which could be the source of the UV excess. Follow-up on this system is warranted to look for close-in white dwarfs, as it was not observed long enough for ADI in this work.

\subsubsection{HD 104018~cc1}

HD~104018~cc1 is a candidate signal $\approx$210~mas (26~au) to the southeast of HD 104018 (TYC~5518-135-1, \textsl{Gaia} source id: 3594365947141542272), a G6-8 subgiant star with RUWE~=~1.14. We performed a KLIP ADI reduction with 5, 10, and 20 modes in $z$ and $i$ band. The candidate signal was observed in KLIP reduced images $z^{\prime}$ and $i^{\prime}$, shown in Figure \ref{fig:comp-images}. It did not scale with wavelength, as the other artifacts did, and the self-subtraction lobes on either side are typical of an ADI-reduced genuine astrophysical signal, so we determined it to be a candidate companion signal. In order to estimate the candidate signal's properties we performed a negative signal injection described in Section \ref{sec: detections} and estimated the separation, pa, and contrast in $z^{\prime}$ and $i^{\prime}$ (it was not observed in $r^{\prime}$). The signal's nature is not well constrained with only $z^{\prime}$ and $i^{\prime}$, but we show its position in an $i^{\prime}$ vs $i^{\prime}$ --$z^{\prime}$ color-magnitude diagram in Figure \ref{fig:comp-cmd}; it clearly falls in the WD sequence, although better photometry is needed to establish its parameters. From our fits to Montreal WD models we determined that it is a relatively cool WD with a H-dominated atmosphere  and \Teff\ constrained to the range 5000~--~11000~K and $\log g$~=~7.5--9.0. The $\chi^2$ surface and best fitting models are shown in Figure \ref{fig:WD-modelfits} (e). Further photometry in bluer bands is necessary to refine the detection and companion properties. Figure \ref{fig:WD-modelfits} shows the model fit results for H-dominated models.

HD~104018 has a PMa measurement from \textsl{Hipparcos}--\textsl{Gaia}, albeit with low significance in the HGCA ($\chi_{\nu}^2 = 3.5$), shown in Figure \ref{fig:PupScc-pma} (bottom).  The orange vertical line marks the separation of HD 104018~cc1, and the horizontal vertical line marks the approximate mass of a white dwarf.  Its location above the sensitivity curve and low HGCA significance indicates HD 104018~cc1 may not be a true astrophysical signal or HD 104018~cc1 is a chance alignment. 

\deleted{\subsubsection{PupS-cc8}}
\deleted{PupS-cc8 is a candidate signal $\approx$140 mas to the north of TYC~169-1942-1. We were unable to determine a spectral type for this star. It is firmly in the giant region of Figure \ref{fig:master target cmd} (bottom) so we cannot use \citealt{PecautMamjek2013MSStarColors} tables, and there is no literature spectral type available. Determination of spectral type for this star is beyond the scope of this work. It is at 415~pc and has an RUWE~=~2.06. 
We performed a KLIP ADI reduction with 5, 10, and 20 modes in $z$ and $i$ band; we did not obtain sufficient rotation for ADI in $r$. Figure \ref{fig:comp-images} shows $z$ and $i$ images reduced with 10 KLIP modes. The candidate signal is marked with a red circle; it is a marginal detection, but did not scale with wavelength and remained present with all KLIP modes. Another spurious signal which does scale with wavelength is marked with a white circle; this is ruled out as a candidate signal. We were unable to characterize the candidate signal using negative signal inject as with cc7 due to the low S/N. Follow-up is necessary with more field rotation and filters to confirm the signal and characterize it. 
We computed contrast curves for this dataset to determine limits for additional companion signals, shown in Figure \ref{fig:nondet-contrastcurves} with a completeness map shown in Figure \ref{fig:nondet-composite-compl-maps} (See section 3.2.1 and 3.3 for descriptions). Our observations rule out additional hydrogen-dominated WD companions beyond from 100--1000~au.
TYC~169-1942-1 has a complete orbital solution in the \textsl{Gaia} DR3 Non-Single Star (NSS) catalog. We computed the predicted location in 2024 March for the two-body solution by converting the Thiele-Innes elements in the \textsl{Gaia} archive to Campbell elements using \cite{Halbwachs2023GaiaAstrBinary} (Appendix A). The NSS solution predicted the 16~$\Mjup$ companion would be at $\rho = 0.43$~mas ($\rho = 0.3$~au) and pa$=216^{\circ}$, which does not correspond to our candidate signal in Section \ref{sec: detections} and is far too close for us to resolve in imaging. }

\movetabledown=0.5in
\begin{longrotatetable}
\setcounter{table}{1}
\begin{deluxetable*}{l|cc|cccccc}
\tablecaption{Companion Detections \label{tab:comp-properties}}
\tablehead{
\colhead{Property} & \multicolumn{2}{c}{TYC 4831-473-1 B} &  \colhead{HD 92588 cc1} & \colhead{BD+20 2007 cc1} & \colhead{BD-00 2082 cc1} & \colhead{TYC 1451-111-1 cc1} & \colhead{HD 108738 cc1} & \colhead{HD 104018 cc1} 
}
\startdata
\hline
\multicolumn{9}{c}{MS Star Host Properties}\\
\hline
TYC & \multicolumn{2}{c|}{4831-473-1} & 4913-1224-1 & 1385-562-1 & 4865-655-1 &  1451-111-1 & 288-976-1 &  5518-135-1  \\
SpT\tablenotemark{a} & \multicolumn{2}{c|}{G2V$^*$} & G9IV$^1$ & G0V$^*$ & F8V$^*$ & F8V$^*$ & G0$^{2}$ & G6/8IV$^{3}$  \\
\deleted{\textsl{Gaia} DR3 ID& \multicolumn{2}{c|}{} &  &  &   &  &  &  \\}
RUWE & \multicolumn{2}{c|}{1.14} & 1.11 & 0.82 & 0.83 &  1.14 & 1.58 & 1.14 \\
Distance\tablenotemark{b} [pc]& \multicolumn{2}{c|}{120.8$^{+0.4}_{-0.2}$} & 37.66$\pm$0.04 & 118.2$\pm$0.3 & 202.7$\pm$0.6  &  222.9$^{+0.9}_{-1.0}$ & 158$\pm$0.8 & 122.3$\pm$0.5\\
\hline
\multicolumn{9}{c}{Companion Properties}\\
\hline
& 2022 & 2024 & & & \\
\hline
Sep (mas) & 908.5$\pm$1.5 & 910.8$\pm$1.3 & 534.9$\pm$0.1 &  953.6$\pm$1.7 & 396.1$\pm$1.8 &  553.7$\pm$0.7 & 1996.0$\pm$1.2 & 230$\pm$9\\
Sep (au) & 110.2$\pm$0.3 & 110.4$\pm$0.3 & 20.2$\pm$0.2 & 113.2$\pm$0.3 & 80.8$\pm$0.5 &  123.7$\pm$0.6 & 318$\pm$1 & 29$\pm$2 \\
P.A. (deg) & 304.1$\pm$0.3& 302.9$\pm$0.6 & 289.0$\pm$0.4 & 16.5$\pm$1.3 & 199.6$\pm$0.5 & 31.6$\pm$0.1 & 254.9$\pm$0.1 & 98.3$\pm$0.3\\
Best-fit model & -- & H dom. WD & MS Star & H dom. WD & H dom. WD & H dom. WD & MS Star & H dom. WD  \\
T$_{\rm{eff}}$ [K] & -- & 8500 -- 17000 & 3100$^{+400}_{-500}$& 6000 -- 20000 &  10500 -- 30000 &  7000 -- 40000 & $<$3500 & 5000 -- 11000\\
$\log g$ & -- & 7.5 -- 8.5 & & 7.0 -- 9.0 & 7.75 -- 9.0  & 7.0 -- 9.0 & -- & 7.5 -- 9.0 \\
$r^\prime$ contrast\tablenotemark{c} & 1.30 [0.06] $\times 10^{-3}$ & 1.36 [0.06] $\times 10^{-3}$ & 2.5 [0.2] $\times 10^{-3}$ & 6.0 [0.8] $\times 10^{-4}$ & 1.5 [0.1] $\times 10^{-3}$ & 1.7 [0.2] $\times 10^{-3}$ & 1.6 [0.6] $\times 10^{-4}$ & --  \\
$r^\prime$ S/N & 23.0 & 23.8 & 11.4 & 7.1 & 14.1 &  9.4 & 2.6 & --\\
$r^\prime$ N$_{\rm{images}}$ & 1162 & 642 & 710 & 822 & 733 & 1101 & 845 & --  \\
$i^\prime$ contrast\tablenotemark{c} & 8.5 [0.4] $\times 10^{-4}$ & 9.9 [0.7] $\times 10^{-4}$ & 3.5 [0.2] $\times 10^{-3}$ & 4.3 [0.4] $\times 10^{-4}$ & 8.4 [0.5] $\times 10^{-4}$  &  1.17 [0.05] $\times 10^{-3}$ & 1.09 [0.06] $\times 10^{-3}$ & 2.2 [0.4] $\times 10^{-4}$ \\
$i^\prime$ S/N & 21.7 & 13.1 & 22.0 & 11.1 & 15.8 &  22.9 & 18.1 & 5.2\\
$i^\prime$ N$_{\rm{images}}$ & 1381 & 705 & 825 & 747 & 738  & 831 & 557 & 869 \\
$z^\prime$ contrast\tablenotemark{c} & -- & 8 [6] $\times 10^{-4}$ & 6.02 [0.09] $\times 10^{-3}$ & 3 [1] $\times 10^{-4}$ & 8 [2] $\times 10^{-4}$ &  9 [2] $\times 10^{-4}$ & 2.32 [0.08] $\times 10^{-3}$ & 1.5 [0.2] $\times 10^{-4}$\\
$z^\prime$ S/N & -- & 5.4 & 38.4 & 2.2 & 3.5 &  3.1 & 28.5 & 5.4 \\
$z^\prime$ N$_{\rm{images}}$ & -- & 706 & 878 & 804 & 734 &  844 & 561 &  708 \\
\enddata
\tablenotetext{a}{Where not available, spectral type is estimated from \citealt{PecautMamjek2013MSStarColors} and \textsl{Gaia} DR3 colors and marked with an asterisk. Spectral Type references: $^{1}$\citealt{Torres2006SACYi}, $^{2}$\citealt{Cannon1993HDcat}}
\tablenotetext{b}{From \citealt{BailerJones2021Distances} catalog}
\tablenotetext{c}{Units Ergs s$^{-1}$ cm$^{-2}$ \AA$^{-1}$; uncertainties in brackets}
\end{deluxetable*}
\end{longrotatetable}

\subsection{Non-Detections}

For targets without a candidate companion detection, we produced contrast curves to estimate sensitivity in a given dataset, and estimated the fraction of H-dominated WDs we would have detected in regions of (mass,physical separation) space for each dataset. In this section we describe how these estimates were produced and the sensitivities for each target observed that did not contain an candidate signal.

\subsubsection{Contrast curves}\label{sec: cont-curves}

For each star without a detected companion, we computed contrast limits for our observation following the method of \cite{Mawet2014} as described in Section 4.4 of \cite{Pearce2022BDI}. To summarize briefly, we performed an injection-recovery analysis by injecting a synthetic signal (made from the PSF reference used in reduction) at a given separation ($r = n\lambda/D$ where $n$ is an integer) and position angle and known contrast into each image in a dataset, then reduced the dataset as before.  We then measured the counts within an aperture of size $\lambda/D$ centered on the injected signal, and the standard deviation of counts within $N-2$ apertures in a ring of radius $r$, where $N = 2\pi r$ and using $N-2$ to omit the apertures on either side of the injected signal to avoid the PSF and ADI self-subtraction lobes, and computed the signal-to-noise ratio (S/N) following Eqn (9) of \cite{Mawet2014}.  We repeated this for all $N$ apertures, took the mean as the S/N for each $r$ at an array of contrasts, then interpolated to find the 2-, 3-, and 5$\sigma$ contrast limits.  Images and contrast curves for each non-detection are shown in Figures \ref{fig:nondet-images} and \ref{fig:nondet-contrastcurves}.

\subsubsection{Survey Completeness}

\begin{figure*}
\centering
\includegraphics[width=0.9\textwidth]{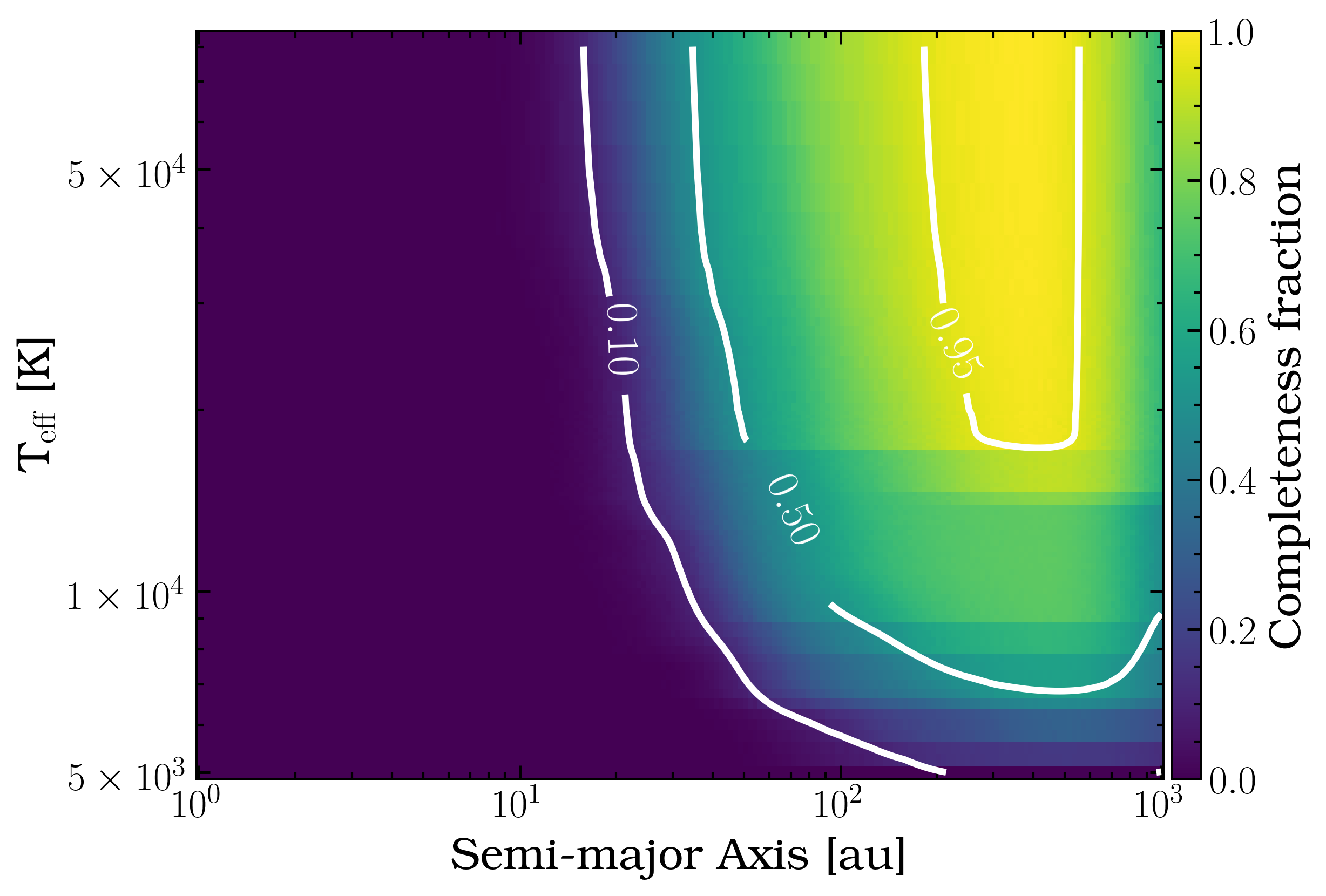}
\caption{\small{Map of completeness to hydrogen-dominated white dwarf companions in our survey to date. The contours and shading mark the fraction of WD companions we would have detected in our survey in regions of (sma,\Teff) space, with 1.0 being all hydrogen-dominated WDs hotter than 5000~K being detected, 0 being none detected. 
}}
\label{fig:Full-completeness-map}
\end{figure*}

To estimate our sensitivity to white dwarf companions, we computed the survey completeness for each ADI and RDI-reduced system and total completeness for the entire survey. To estimate completeness we simulated 100 white dwarf companions over a grid uniform in log(semi-major axis)~$\in$~[0, 3] and the H-dominated white dwarf model temperatures in the \citetalias{koesterWhiteDwarfSpectra2010} models (T$_{\rm{eff}}$ = 5000--20000~K spaced by 250~K, 20000--30000~K spaced by 1000K, 30000--40000~K spaced by 2000~K, and 40000--80000~K spaced by 10000~K). For each MS star we computed an array of MS-WD contrast from the \citetalias{koesterWhiteDwarfSpectra2010} models. We generated 5$\times$10$^{3}$ simulated companions for each (sma,contrast) grid point, randomly assigned orbital parameters from priors\footnote{eccentricity (e): P(e) = 2.1 - 2.2$\times$e, e $\in$ [0,0.95], following \citealt{nielsen_gemini_2019}; inclination (i): $\cos$(i) $\in$ Unif[-1,1]; argument of periastron ($\omega$): $\omega \in$ Unif[0,2$\pi$]; mean anomaly (M): M $\in$ Unif[0,2$\pi$]; since contrast curves are one-dimensional we did not simulate longitude of nodes}, and computed projected separation.  A companion was considered detectable if it fell above the 2$\sigma$ contrast curve and undetectable if below the curve, inside the inner limit of the curve, or outside the detector. The completeness is the fraction of simulated companions at each grid point that would have been detectable at at least S/N = 2; 1.0 corresponds to every simulated companion in a bin being detected, 0.0 corresponds to no companions being detected. Figure \ref{fig:nondet-composite-compl-maps} shows the completeness for each observed system sensitive to close-in companions (ADI or RDI reduced systems). Figure \ref{fig:Full-completeness-map} shows the combined completeness map for the entire survey to date. Of the targets observed in this work, we are nearly complete to WD companions hotter than 20000~K between 100--1000~au, and have some fractional completeness down to 50~au. These limits can be improved by longer ADI field rotation times and the use of a coronagraph.

\subsubsection{Detection limits}
\textbf{\textit{TYC~1262-1500-1:}} an early F star at 452~pc with RUWE~=~1.37 (\textsl{Gaia} source id: 52921148211018368). We reduced the data using both classical and KLIP ADI with 10 deg of field rotation and did not detect any candidate signals. Our observations rule out H-dominated WDs hotter than 20000~K from $\approx$300--1000~au. The elevated RUWE hints at a large-luminosity-ratio companion at a closer radius. Longer time baseline ADI observations will improve detection limits for this system.

\textbf{\textit{HD 87147:}} (TYC 5480-589-1), a K0III giant star at 630 pc with RUWE = 1.42 (\textsl{Gaia} source id: 3766625330057013248). We reduced the data using both KLIP and classical ADI and did not detect any candidate companion signals. \deleted{White circles in Figure \ref{fig:nondet-images} mark speckles that scale with wavelength and are not candidate signals.} Our observations rule out H-dominated WDs hotter than 6000~K from 60--1000~au. The high RUWE value hints to a large-luminosity-ratio companion being closer to the star than we were able to obtain with our observations.  Better coronagraphs in the near-term and the next generation of large ground-based telescopes could enable detection of the source of the UV excess and astrometric perturbation.

\textbf{\textit{TYC~5512-916-1:}} a giant branch star at 415~pc with RUWE~=~3.03. We reduced the data using both KLIP and classsical ADI and did not detect any candidate companion signals. Our observations rule out H-dominated WD companions from 30--1000~au. This star has one of the highest RUWE values in our sample, suggesting that the large-luminosity-ratio companion is close to the host star.  Longer time-baseline ADI may improve contrast limits as close separations, and better coronagraphs and larger telescopes could enable detection.

\textbf{\textit{CD-28 10038:}} (TYC~6712-1511-1), a giant star at 174~pc with RUWE~=~1.09 (\textsl{Gaia} source id: 6183075365029951488). We were unable to attain sufficient field rotation for ADI for this system due to observing conditions.  We reduced these data using classical reference differential imaging (RDI) using TYC~5512-916-1 images as stellar PSF reference, which was obtained earlier the same night under the same conditions. For each filter, we created a PSF reference as the median TYC~5512-916-1 image in the same filter, scaled the reference to the peak flux in CD-28 10038, and reduced the images from there in the same manner as classical ADI.  This was not optimal, as reflected in our contrast limits in Figure \ref{fig:nondet-contrastcurves}, and should be revisited with longer time-baseline ADI.  We did not detect any candidate companion signals.  Our observations rule out H-dominated WDs hotter than 20000~K from 30--1000~au

\textbf{\textit{HD 109439:}} (TYC~877-681-1), a G5 star at 260 pc and RUWE~=~1.00 (\textsl{Gaia} source id: 3904521719021034880). Seeing was variable during our observations and\deleted{and we were limited in high-quality images suitable for reduction, consequently} we attained less than ten degrees of rotation. We reduced these images using HD 104018, observed on the same night, as a reference for RDI. Our contrast curves in Figure \ref{fig:nondet-contrastcurves} achieve higher contrast than for CD-28 10038, but not as high as our ADI-reduced datasets. Our observations rule out WDs hotter than $\approx$8000~K from 30--1000~au. This target will be re-observed in future observing campaigns. \added{It has both \textsl{Hipparcos} and \textsl{Gaia} astrometry with HGCA $\chi_{\nu}^2 = 180$, and Figure \ref{fig:nondet-pma-plots} (b) shows the mass of an object that would cause the observed acceleration as a function of separation. Our contrast curves rule out companions greater than 0.32\Msun\ beyond 0.09\arcsec. \citetalias{Ren2020AFGKSample} predicted a WD companion with \Teff~=~11700$^{+150}_{-140}$~K; our contrast curves rule out a companion of that temperature within 50--1000~au with at least 50\% probability, shown as the orange line in Figure \ref{fig:nondet-composite-compl-maps} (e).}

\textbf{\textit{TYC~169-1942-1:}} a giant star at 415~pc and has an RUWE~=~2.06 (\textsl{Gaia} source id: 3136564420989284096). \added{We observed this star in 2022~Dec and again in 2024~Mar, reducing each epoch with KLIP ADI, with no candidate signal in either epoch.} Figure \ref{fig:nondet-images} shows a 2024 $i^{\prime}$ image reduced with 20 KLIP modes. 
\deleted{We computed contrast curves for this dataset to determine limits for additional companion signals, shown in Figure \ref{fig:nondet-contrastcurves} with a completeness map shown in Figure \ref{fig:nondet-composite-compl-maps} (See section 3.2.1 and 3.3 for descriptions). Our observations rule out additional hydrogen-dominated WD companions from 100--1000~au.} TYC~169-1942-1 has a complete orbital solution in the \textsl{Gaia} DR3 Non-Single Star (NSS) catalog. We computed the predicted location in 2024 March for the two-body solution by converting the Thiele-Innes elements in the \textsl{Gaia} archive to Campbell elements using \cite{Halbwachs2023GaiaAstrBinary} (Appendix A). The NSS solution predicted a 16~$\Mjup$ companion would be at $\rho = 0.43$~mas ($\rho = 0.3$~au) and pa$=216^{\circ}$, which \deleted{does not correspond to our candidate signal in Section \ref{sec: detections} and }is far too close for us to resolve in imaging. \added{\citetalias{Ren2020AFGKSample} predicted a WD companion with \Teff~=~30000$^{+2500}_{-2300}$~K; our contrast curves rule out a companion of that temperature within 90--1000~au with at least 50\% probability, shown as the orange line in Figure \ref{fig:nondet-composite-compl-maps} (f).}

\textbf{\textit{TYC~1447-1616-1:}}, an F8 star at 400 pc with RUWE = 1.03 (\textsl{Gaia} source id: 3953075946345237632). We were able to get 17$^{\circ}$ of rotation during our observations and did not find any candidate signals in our KLIP ADI reduced images. Our observations rule out WDs hotter than $\approx$15000~K from 100--1000~au

\textbf{\textit{HD 146740:}} (TYC~368-1591-1), a K0III star at 220~pc with RUWE~=~1.51 (\textsl{Gaia} source id: 4408906185595207424). Observing conditions were not ideal (seeing $\approx$1.1\arcsec, which affected the AO correction), so future imaging could get to deeper contrasts. Our observations rule out WD's hotter than 6000~K from 60--1000~au. It has a complete orbital solution in \textsl{Gaia} which predicts a 4.7~\Mjup\ planet at 0.9~mas (0.2~au), much too close for us to resolve in imaging. It was also observed in \textsl{Hipparcos} and has an observed acceleration between \textsl{Hipparcos} and \textsl{Gaia} astrometry with HGCA $\chi_{\nu}^2 = 105$. Our imaging rules out objects that could be causing the observed acceleration at 0.4\Msun\ beyond 160~mas, shown in Figure \ref{fig:nondet-pma-plots} (a). \added{\citetalias{Ren2020AFGKSample} predicted a WD companion with \Teff~=~25000$^{+2000}_{-1900}$~K; our contrast curves rule out a companion of that temperature within 45--1000~au with at least 50\% probability, shown as the orange line in Figure \ref{fig:nondet-composite-compl-maps} (h).}

\added{

\textbf{\textit{HD 7918:}} (TYC 26-39-1), a G5 star at 295~pc with RUWE~=~1.15 (\textsl{Gaia} source id: 2578318384467341056). We obtained 18$^{\circ}$ rotation and reduced the data using KLIP ADI but did not find any candidate signals. Our imaging rules out H-dominated WDs from 60--1000~au. \added{It also displays a significant proper motion anomaly between \textsl{Hipparcos} and \textsl{Gaia} with HGCA $\chi_{\nu}^2 = 640$; our imaging rules out companions $>$0.95\Msun\ beyond 0.09\arcsec. \citetalias{Ren2020AFGKSample} predicted a WD companion with \Teff~=~15000$^{+1000}_{-900}$~K; our contrast curves rule out a companion of that temperature within 25--1000~au with at least 50\% probability, shown as the orange line in Figure \ref{fig:nondet-composite-compl-maps} (i).}

\textbf{\textit{CD-27 191:}} (TYC 6423-1892-1), a giant star at 251~pc with RUWE~=~1.00 (\textsl{Gaia} source id: 2343530870495486464). We obtained over 100$^{\circ}$ rotation in $i^{\prime}$ and $z^{\prime}$ but did not find a candidate signal in KLIP reduced images. Our detection limits rule out WDs hotter than 6000~K at 50--1000~au. 

\textbf{\textit{HD 680:}} (TYC 5-436-1), a late-G/early-K type giant star at 270~pc with RUWE~=~1.36 (\textsl{Gaia} source id: 2740897190273065344). We did not detect any candidate signals in KLIP-reduced ADI images. Our imaging rules out WDs hotter than 8000~K from 50--1000~ au. \added{It also displays a signficant proper motion anomaly between \textsl{Hipparcos} and \textsl{Gaia} with HGCA $\chi^2 = 390$, and our imaging rules out companions $>$1\Msun\ beyond 0.09\arcsec. \citetalias{Ren2020AFGKSample} predicted a WD companion with \Teff~=~49000$^{+4700}_{-4800}$~K; our contrast curves rule out a companion of that temperature within 30--1000~au with at least 50\% probability, shown as the orange line in Figure \ref{fig:nondet-composite-compl-maps} (k).}

\textbf{\textit{BD-01 220:}} (TYC 4685-1113-1), a giant star at 600~pc with RUWE~=~2.38 (\textsl{Gaia} source id: 2508886355477047296). Our detection limits rule out H-dominated WDs hotter than 5000~K at 80--1000au. The high RUWE value hints the UV-excess companion may be closer than we can reach with our imaging. \added{\citetalias{Ren2020AFGKSample} predicted a WD companion with \Teff~=~41000$^{+3700}_{-3600}$~K; our contrast curves rule out a companion of that temperature within 35--1000~au with at least 50\% probability, shown as the orange line in Figure \ref{fig:nondet-composite-compl-maps} (l).}
}

\section{Discussion}

In this section we discuss how our survey fits in the context of other SLS surveys and future prospects for the Pup Search program.

\begin{figure*}
\centering
\includegraphics[width=0.9\textwidth]{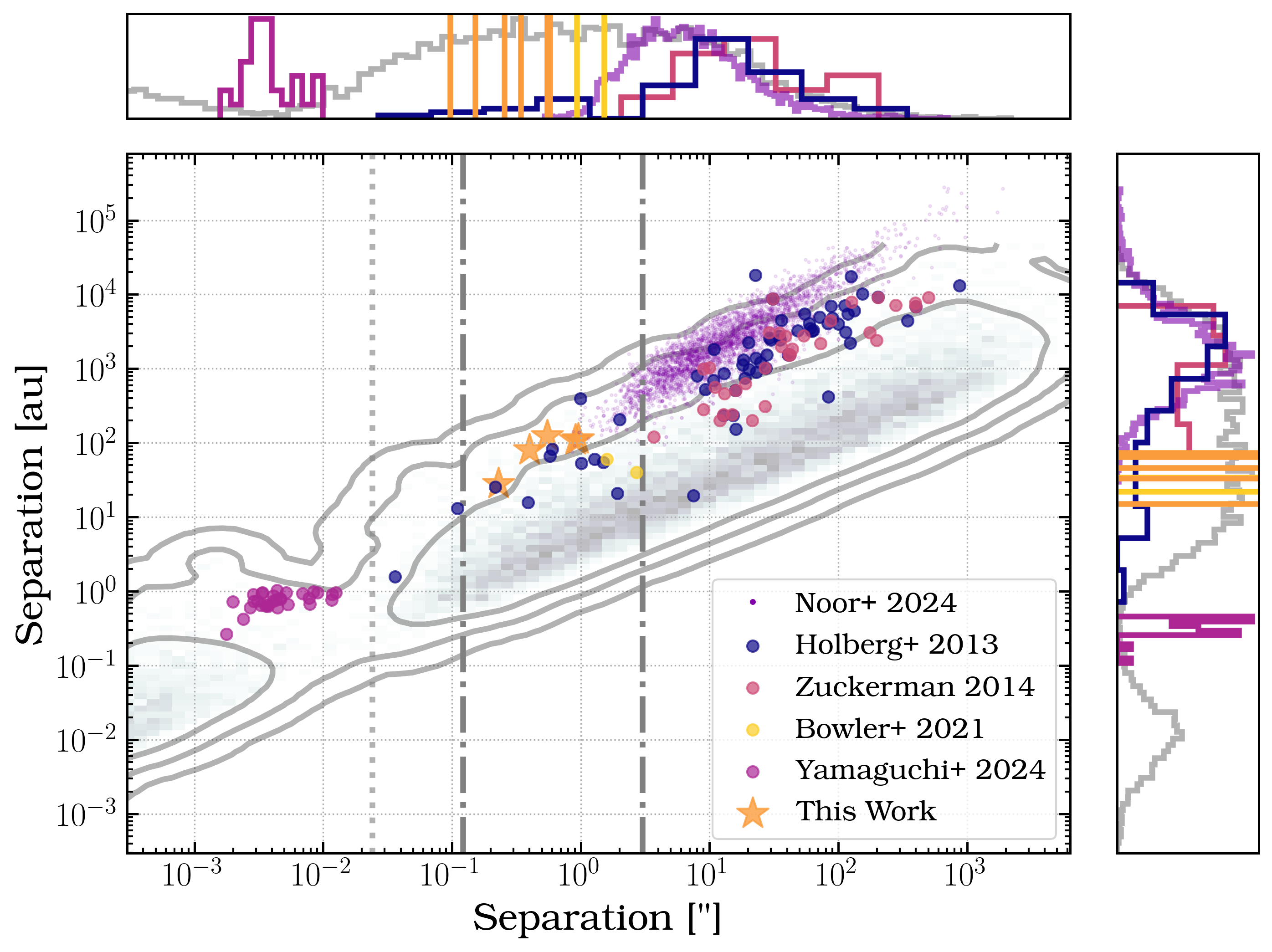}
\caption{\small{Comparison of our work to previous SLS publications in angular separation vs physical separation. 
Confirmed and candidate companions in this work (orange stars) are plotted against known SLS from \citealt{zuckermanOccurrenceWideorbitPlanets2014} (red), \citetalias{Holberg2013} (and references there-in; blue), WDMS systems in \textsl{Gaia} DR3 from \citealt{Noor2024WDPollutionInWideBinaries} (purple), wide post-common envelope binaries in \textsl{Gaia} from \citealt{Yamaguchi2024WDMS-Gaia} (magenta), and two SLS detected via astrometric acceleration and AO imaging in \citealt{Bowler2021WD-DynMasses} (yellow). The grey density gives an adaptation of \citealt{Willems2004DetechedWDMS} Figure 10 (right) to our parameter space, with contours marking 1--, 2--, and 3--$\sigma$ confidence intervals. We illustrate the approximate limits of MagAO-X search space between 120 mas (5 $\lambda/D$ in $i'$, a conservative estimate of smallest angular resolution achievable) and 3'' (half the detector FOV) by grey dash-dot lines; the light grey dotted line marks the expected capabilities of fully optimized coronagraphs on MagAO-X with $\approx1\lambda$/D inner working angle.} Our survey can bridge a gap between SLS direct imaging and detection via other means.
}
\label{fig:pups-compared-to-others}
\end{figure*}

\subsection{Our survey compared to previous surveys}

\textbf{\textit{ExAO probes tighter angular and physical regimes.}} Our survey is able to resolve two signals at closer angular separations and smaller WD--MS physical separations than most previous SLS detections. 
Figure \ref{fig:pups-compared-to-others} shows our new SLS in angular separation compared to physical separation, and compares to other WD+MS star surveys: \citet{zuckermanOccurrenceWideorbitPlanets2014} (red), \citetalias{Holberg2013} (blue), and \citet{Noor2024WDPollutionInWideBinaries} (magenta), which all looked at non-interacting SLS as discussed in Section \ref{sec:intro}. For comparison we also include \citet{Yamaguchi2024WDMS-Gaia}, which examined the orbits of post-common envelope binaries or mass transfer WD+MS systems with spectroscopic orbital solutions in \textsl{Gaia} DR3. The approximate detection region for MagAO-X is marked with dash-dotted grey lines and indicate the regime we are sensitive to (from $5\lambda/D$ at 0.8$\mu$m to 3''). Approximately 30\% of \citetalias{Holberg2013} and 6\% of \citet{Noor2024WDPollutionInWideBinaries} objects fall within that regime. Additionally, we include an inner resolution limit predicted to be achieved by ongoing coronagraph updates to MagAO-X (Sec \ref{sec: future}). We overlaid density and contours adapted from Figure 10 (right) from \citet[][hereafter W04]{Willems2004DetechedWDMS} to this parameter space. That figure gives a distribution of orbital periods for all types of WD+MS systems at the start of the WD phase. The distribution is bimodal -- as the progenitor evolves from the giant branch into the WD phase, the orbit either contracts into a common-envelope phase (\citetalias{Willems2004DetechedWDMS} channels 1--6) or expands as the progenitor loses mass (\citetalias{Willems2004DetechedWDMS} channel 7, non-interacting systems). To create the contours we drew a Monte Carlo sample of orbital periods from the distribution in \citetalias{Willems2004DetechedWDMS} Fig. 10 (right), drew WD and MS masses from the distributions in \citetalias{Willems2004DetechedWDMS} Fig. 10 (left, center respectively), and drew distances from a uniform distribution spanning distances in the \citetalias{Holberg2013} sample. We drew orbital inclination from a uniform $\cos i$ distribution and then computed projected separation in au and arcseconds.

Our WD detections in this work are plotted in orange. We see that our detections are at closer angular and physical separations than most known SLS, in a regime where non-interacting SLS systems are expected to reside. This region is not well probed by other methods.

\begin{figure*}
\centering
\includegraphics[width=0.9\textwidth]{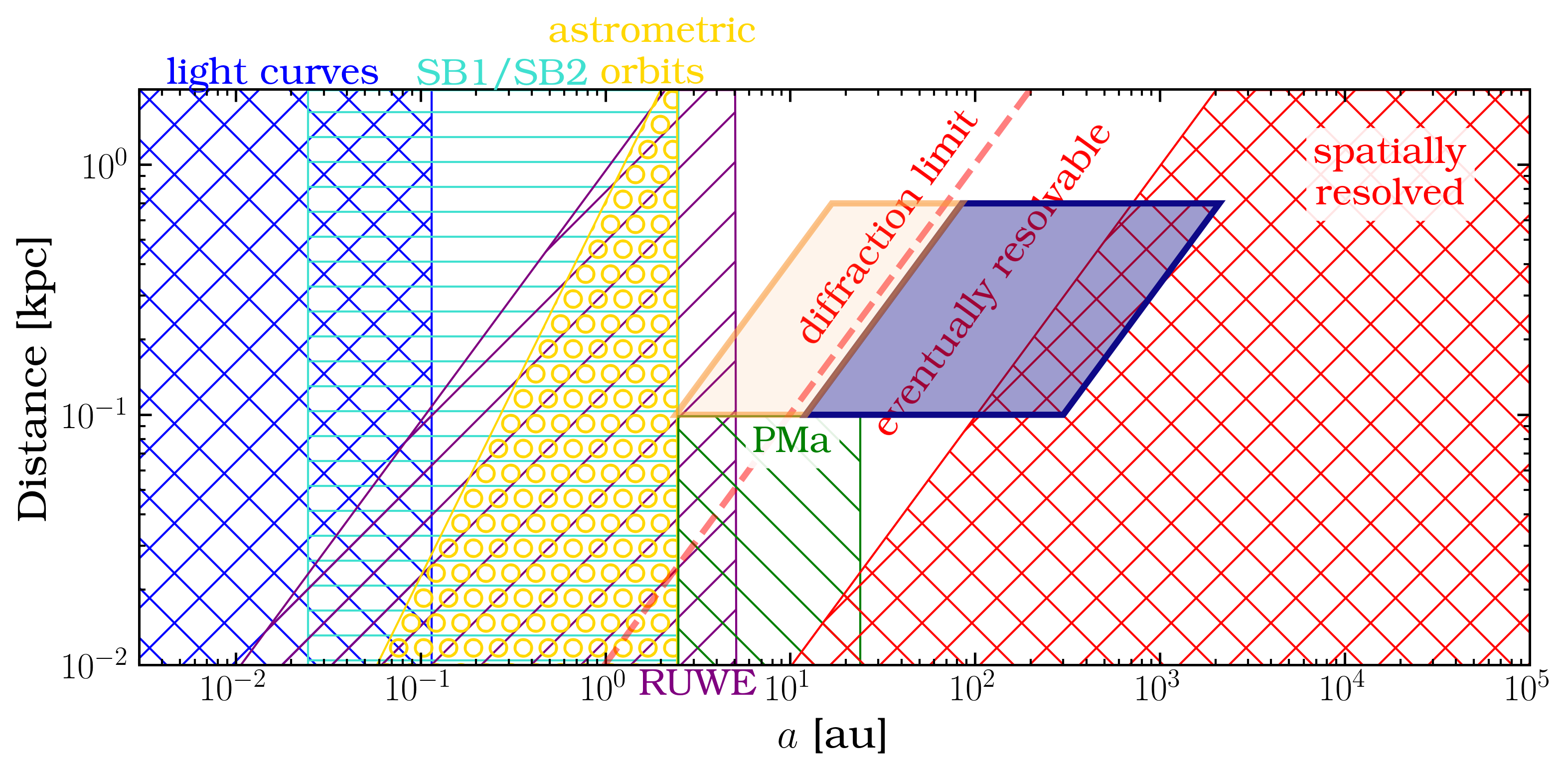}
\caption{\small{Reproduction of Figure 1 from \citet{ElBadry2024GaiaBinaries} showing the sensitivity regimes of various multiplicity metrics for a two-solar mass binary. The purple filled region indicates the regime of the targets in our survey, at $\approx$100--700~pc; the inner and outer au ranges of the purple region are the inner and outer angular sensitivities from Figure \ref{fig:pups-compared-to-others} converted to physical separation as a function of distance. The orange filled region shows the sensitivity regime for fully optimized coronagraphs} (see Section \ref{sec: future})
They fall outside the regime for which RUWE and PMa are most sensitive, and even fall inside the regime eventually resolvable by \textsl{Gaia}. Our detected companions are too faint to be resolved in \textsl{Gaia} even if they do fall within the spatially resolvable region. That these metrics are useful but incomplete for assessing WD multiplicity, and our survey will probe a regime not probed by any other multiplicity metric.
}
\label{fig:elbadry-ruwe}
\end{figure*}

\textbf{\textit{Our survey probes a regime not accessible to }\textrm{Gaia}.} In Figure \ref{fig:elbadry-ruwe} we reproduced Figure 1 from \citet{ElBadry2024GaiaBinaries} showing the sensitivity regime for various \textsl{Gaia} multiplicity metrics (for a pair of solar mass stars). We have overplotted the filled purple region indicating the regime where our survey is sensitive. We used the inner and outer angular resolution limits from Figure \ref{fig:pups-compared-to-others} and distances of 100--700~pc, the distances of the targets in the Pup Search. The filled orange region represents the sensitivity regime extrapolated to future coronagraph improvements (Sec \ref{sec: future}). We are sensitive to regions not covered by \textsl{Gaia} astrometric orbits, RUWE, or proper motion anomaly. Our survey has the potential to be sensitive to regions not detectable even to future \textsl{Gaia} spatial resolution. Additionally, the companions detected here are too faint to be detected by \textsl{Gaia} even in the spatially resolved regime, as the \textsl{Gaia} magnitude limit is $G\lesssim20$~mag \citep{GaiaCollaboration2016}. Our survey probes companions too faint to be resolved yet too wide to affect multiplicity metrics, a regime not probed by any other metric.

\textbf{\textit{Orbits of SLS in the 10--100~au regime warrant study.}} Detection and orbital characterization of SLS in this regime will help constrain occurrence rates and simulations of the effect of the companion on planetesimals around the WD. For example, if a large fraction of SLS in the 10--100~au regime prove to be on highly-eccentric orbits with larger semi-major axes, we could infer that objects in this group may be undergoing eccentricity evolution into a high-e state as predicted by \citet{Stegmann2024WideBinariesGalacticTide}. Simulations of planetstimal perturbation in these states are warranted \citep[e.g.][]{Stephan2017ThrowingIcebergs}. If all are on nearly-circular orbits, then perturbation by external perturbers with subsequent eccentricity evolution is less likely \citep{Stegmann2024WideBinariesGalacticTide} and pollution is less likely to be influenced by the companion. Previous surveys of more widely separated pairs (those in Figure \ref{fig:pups-compared-to-others}) concluded no evidence for companion influence in pollution; this survey provides a slightly different SLS population for which to test this mechanism.

\deleted{
\subsection{Towards probes of companion influence on pollution through orbit studies}

While the previous studies discussed above have looked at the influence of binarity on WD pollution from a statistical perspective, none to date have examined the orbits of polluted vs non-polluted non-interacting Sirius-Like-Systems observationally. Wide (100's -- 1000's au) stellar binary orbits evolve on Gyr timescales due to the influence of the Galactic gravitational potential and stellar flybys into (depending on initial conditions) oscillating high and low eccentricity states \citep{Kaib2013GalacticPerturbations, bonsorWideBinaryTrigger2015, correa-otto_galactic_2017, Stephan2017ThrowingIcebergs} with close periastron passages \citep{Heisler1986GalacticTide, Collins2008ImpulsiveEncounters, Modak2023GalacticTidesEccentricity}
even reaching eccentricities as high as 0.99 \citep{Stegmann2024WideBinariesGalacticTide}. Binaries with semi-major axis from 1000~--~10,000~au are most susceptible to high-e oscillations without disruption on timescales less than a Hubble time \citep{Stegmann2024WideBinariesGalacticTide}. WDMS systems are especially vulnerable as the WD progenitor looses mass and the orbit expands, and planetesimals can be scattered onto star-grazing orbits, especially if a planet is present \citep{bonsorDynamicalEffectsStellar2011, debesLinkPlanetarySystems2012}. 
During high-e periastron passages, planetesimals will be perturbed if the binary passes interior to $\approx$500 au \citep{bonsorWideBinaryTrigger2015}. Approximately 20\% of wide binaries are estimated to have orbital parameters in this range to drive WD pollution \citep{bonsorWideBinaryTrigger2015}. The binary pollution mechanism may still play a role in pollution for individual systems.

Orbital studies of polluted and non-polluted SLS systems are warranted. Many separations in this and previous SLS surveys are in the range of susceptibility to eccentricity perturbation without disruption; a high-e binary pollution mechanism may still be at play for this population, detectable via an observable trend in eccentricity for polluted SLS systems compared to non-polluted.

A long term goal of the Pup Search is to obtain time series astrometry of new and known polluted and non-polluted SLS systems resolvable to direct imaging. This will require a long-term effort since short-orbit-arc fitting ($\lesssim$50\% of the orbital period) artificially inflates the prevalence of high-eccentricity orbits in the posterior, which is the main parameter of interest for this science case \citep{Ferrer-Chavez2021OFTIBiases}. Long astrometric time baseline and precision radial velocities of both components will be important. Additionally future \textsl{Gaia} data releases containing time series astrometric data, more sources with spectra and radial velocities, and even some SLS in the 100-1000~au separation range with determined orbits will contribute to these studies (while methods currently exist to use \textsl{Gaia} data for wide binary orbit fitting [e.g. \citealt{Pearce2020}], these data are too loosely constraining and will exhibit biased eccentricity results). Proper motion anomaly can also help constrain SLS orbits where available. \citet{Zhang2023DynMassesof6SLS} determined dynamical masses and orbital parameters of six confirmed and one candidate SLS using PMa, however the majority of our target list does not have a \textsl{Hipparcos -- Gaia} proper motion measurement, so careful observation and analysis will be required to constrain orbital parameters.}

\subsection{Revisiting the ``missing white dwarf problem"}
\citet{Holberg2016-25pcLocalWDPop} predicted that as many as 100 white dwarfs were ``missing'' from the 25~pc local volume. \citet{Katz2014LuminosityFunction} Figure 1 displays a histogram of number of WDs within 20~pc in bins of absolute $V$ magnitude for single WDs, SLS systems from \citet{Holberg2008NewLookatLocalWDPop}, and the number predicted from theory. The single WD population matches the prediction well, but the SLS population falls below the prediction for WDs fainter than 11th absolute $V$ magnitude, implying WD companions are being missed, possibly due to bright hot MS companions. With new \textsl{Gaia} data we can revisit that plot.

Since the sea-change provided by \textsl{Gaia}, the local population of single WDs with $G<20$~mag and WDs in multiples in a 100~pc volume is essentially complete \citep{GentileFusillo2023CatalogOfGaiaWD, RebassaMansergas2021GaiaWDMS, Jimenez-Esteban2023100pcWDpop}, with $>$97\% completeness inside 40~pc \citep{OBrien2024-40pcGaiaSample}. \citet{Jimenez-Esteban2023100pcWDpop} determined that white dwarfs resolved in \textsl{Gaia} are 100\% complete for $G_{\rm{BP}} - G_{\rm{RP}} < 0$, $>$90\% for $G_{\rm{BP}} - G_{\rm{RP}} < 0.86$, and decreasing to 70\% at the reddest end. WDs are missing from \textsl{Gaia} due to 1) confusion with the Galactic plane ($\approx1\%$), 2) double degenerate systems counted as single WDs ($\approx1-3\%$), and 3) Unresolved WD+AFGK systems ($<8\%$, \citealt{Holberg2013}). In actuality all but four\footnote{HD 27786/56 Per Ab, HD 149499/HD 149499 B, HD 202109/$\zeta$ Cyg B, BD -7 5906/HD 217411 B} of the \citet{Holberg2013} systems closer than 100~pc are resolved in \textsl{Gaia}, so the actual missing percentage in the \citet{Jimenez-Esteban2023100pcWDpop} catalog is much lower. All of the \citet{zuckermanOccurrenceWideorbitPlanets2014} systems are resolved in \textsl{Gaia} and so are contained in the \citet{Jimenez-Esteban2023100pcWDpop} population. \citet{RebassaMansergas2021GaiaWDMS} used \textsl{Gaia} colors to identify sources falling between the WD and main sequences to find unresolved SLS systems. This method identifies mainly WD companions to M and K stars where the WD dominates the SED. They estimate from population synthesis that their sample represents 9\% of the underlying SLS population, with the other 91\% being located within the main sequence as the MS star dominates the SED. \citet{Nayak2024HuntingDown} identified 93 SLS candidates within 100~pc using \textsl{Gaia} DR3 and \textsl{GALEX} GR6/7 UV measurements, 80 of which are newly identified. They were identified using SED fitting and so remain unresolved.

\begin{figure*}
\centering
\includegraphics[width=0.9\textwidth]{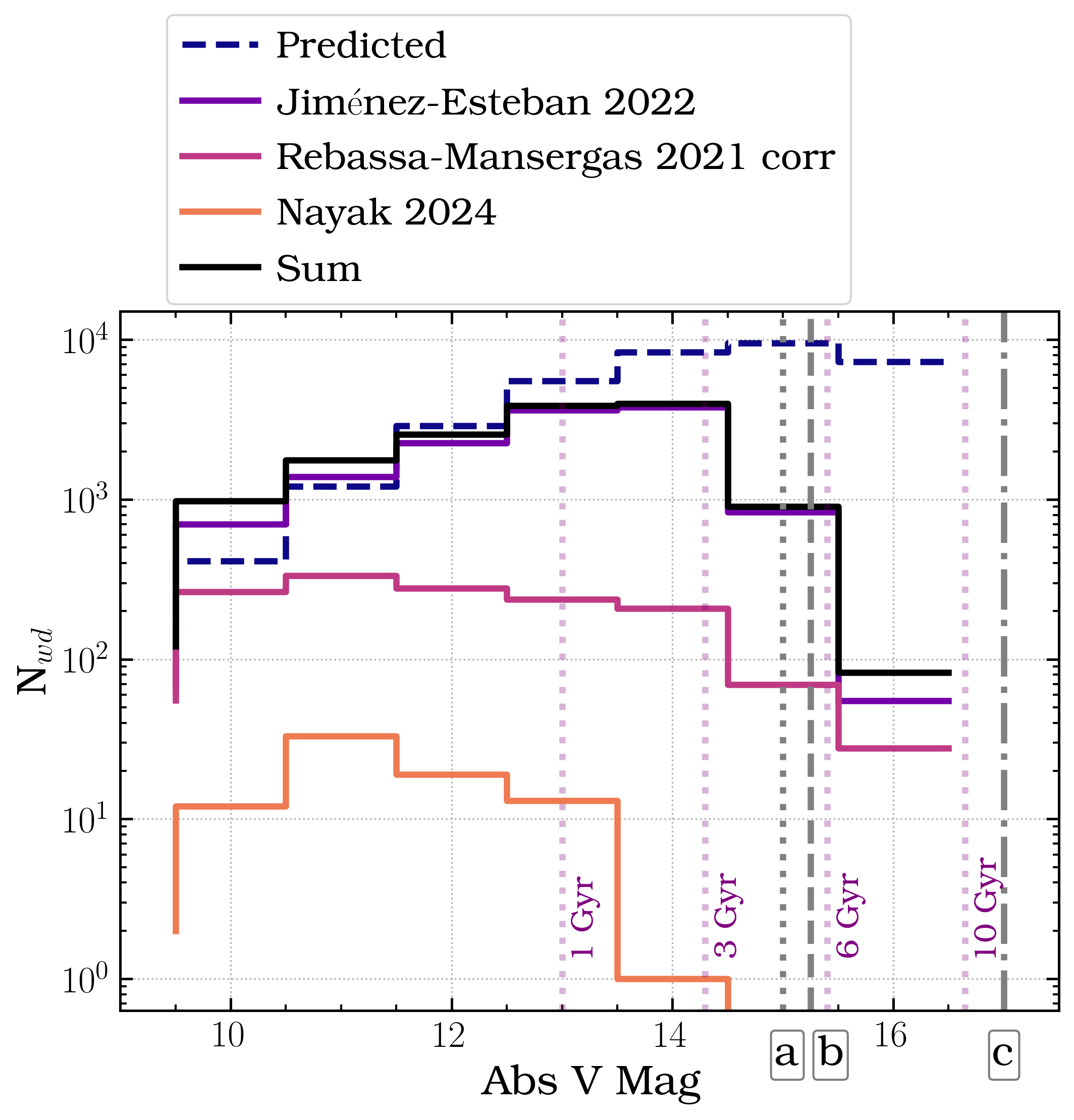}
\caption{\small{Number of WDs expected and observed within 100~pc local volume. The dashed blue line is the number of WDs predicted by theory (see text for details); the purple line is the 100~pc \textsl{Gaia} sample from \citealt{Jimenez-Esteban2023100pcWDpop}; the magenta line is the 100~pc SLS sample from \citealt{RebassaMansergas2021GaiaWDMS}, corrected for the 80\% completeness of their sample and the 9\% completeness of their sample from the total underlying SLS population; the orange line is the new SLS from \citealt{Nayak2024HuntingDown}; and the black line is the sum. We did not include SLS surveys for which the objects are resolved in \textsl{Gaia}, since they would be captured in the \citet{Jimenez-Esteban2023100pcWDpop} catalog. The dotted grey line (a) shows the approximate detection limit for \textsl{Gaia} for a WD at 100~pc, assuming a detection limit of $G = 20$ mag; the dashed grey line (b) shows the detection limit at the background limit of the deepest observations in this work, a flux contrast of approximately 5$\times10^{-5}$, to a G0 type star at 100 pc; the dot-dashed grey line (c) shows the background limit if we are able to achieve a 5$\times$ deeper contrast. The purple dotted lines mark the WD absolute V-band magnitudes as a function of cooling age (Eqn \ref{eq: cooling age}). The faintest white dwarfs are still being missed, however the number of observed WDs from 11--14th magnitude much more closely matches theory than in \citet{Katz2014LuminosityFunction}. Our survey is already probing all but the oldest WDs in the background-limited regime.}
}
\label{fig:katz-redone}
\end{figure*}

Figure \ref{fig:katz-redone} shows the number of WDs in the 100~pc volume as a function of absolute $V$ mag. The dashed blue line shows the number of WDs in bins of absolute $V$ magnitude predicted by theory. We used \citet{Katz2014LuminosityFunction} Eqns 1 and 2 to produce that line:
\begin{equation}
    N \approx \dot n_{\rm{WD}} \left[ t_{\rm{cool}}(M_{V,2}) - t_{\rm{cool}}(M_{V,1}) \right]
\end{equation}
\begin{equation}\label{eq: cooling age}
    \log_{10}(t_{\rm{cool}}/yr) = -0.04M_V^2 + 1.46M_V - 3.22
\end{equation}
where $\dot n_{\rm{WD}}$ is the WD formation rate with $\dot n_{\rm{WD}} = 0.7\times10^{-12}\; \rm{pc}^{-3} \;\rm{yr}^{-1}$, $t_{\rm{cool}}(M_{V})$ is the cooling age for a WD with absolute $V$ mag $M_V$. The purple line is the 100~pc \textsl{Gaia} sample from \citealt{Jimenez-Esteban2023100pcWDpop}, the magenta line is the 100~pc SLS sample from \citealt{RebassaMansergas2021GaiaWDMS}, corrected to 100\% completeness of the underlying SLS population, the orange line is the new SLS from \citealt{Nayak2024HuntingDown}, and the black line is the sum. The sum follows the prediction closely except at the faintest end where WD detection is challenging for the coolest and faintest WDs. The number of observed WDs much more closely matches prediction from 11-14th magnitudes than in \citet{Katz2014LuminosityFunction}.

Only five of the current batch of Pup Search targets are within 100~pc, so our discovery survey is not poised to make much of a dent in that particular parameter space. Assuming a detection limit contrast of 5$\times10^{-5}$ to a G0 star and the same detection rate as in this paper (6/18), we estimate this set of targets will return 28$\pm$5 new SLS. As part of this work we will continue to compile all SLS discoveries, which our results will add to, in our efforts to examine the population as a whole.

\subsection{Future work}\label{sec: future}
In this work we have reported the first results of the Pup Search with the MagAO-X instrument. \added{The photometry of our detected WD companions generally favors cooler temperatures than predicted (where available) in \citetalias{Ren2020AFGKSample}. In addition to imaging the remainder of our target list, the next phase of the Pup Search will include high resolution spectroscopy of our newly-detected companions to determine mass, \Teff, and spectral features of both WD and MS companions.}

Additionally, MagAO-X is an active technology development platform for the next generation of ground-based high-contrast imaging with 30--m class telescopes. The ultimate science goal of MagAO-X is to directly image nearby exoplanets in reflected light \citep{Males2022MagAOX}, requiring advancements in angular resolution and contrast capabilities, driving AO and coronagraph technology development. 
MagAO-X technology development efforts are ongoing as of this publication, and will affect the impact this survey will have on this science case in the future. As shown in Figure \ref{fig:katz-redone}, our current contrast limit probes all but the oldest white dwarf companions in the background-limited regime. High contrast as much as $10\times$ closer will enable this survey to bridge the gap between SLS direct detections and detections via other means such as RV and spectroscopically (Figure \ref{fig:pups-compared-to-others}), and probe regimes inaccessible to other methods (Figure \ref{fig:elbadry-ruwe}).

\section{Conclusion}

We have introduced the ExAO Pup Search, a survey using the tools of extreme AO instrumentation to probe the influence of non-interacting stellar companions to white dwarf stars. We have conducted the first Pup Search observations of 18 target stars and have detected one new confirmed Sirius-Like System, TYC~4831-473-1~B, a H-dominated white dwarf at 110~au (900~mas) from the Sun-like star TYC~4831-473-1. We detected four additional candidate SLS: confirmed white dwarf objects close to AFGK stars which we will be following up with additional observations to confirm they are bound companions. We detected one likely unresolved WD+M-dwarf binary candidate companion 500~mas from the G9IV star HD~92588, and one main sequence companion to HD~108738. Several of our targets without detections had low contrast limits due to observational challenges and merit followup with more field rotation for adequate ADI reduction. Our target list is drawn from a high-quality sample of stars with UV excess and radial velocity trends, so non-detections are likely due to the companion being too close to the star to resolve rather than there not being a companion at all. Given the same detection rate, we predict the Pup Search will yield 28$\pm$5 new Sirius-Like systems. Our survey is sensitive to regimes not currently or previously probed by other methods.\deleted{The goal of the Pup Search is to contribute to the discussion of the influence of wide companions in driving white dwarf pollution by probing regimes inaccessible to previous surveys, and to to probe orbital parameters of polluted and non-polluted systems. }Future work will continue to use the (ever-improving) tools of ExAO and high resolution spectroscopy towards detection and characterization of new SLS.

\appendix{}

\begin{figure*}
\centering
\includegraphics[width=0.98\textwidth]{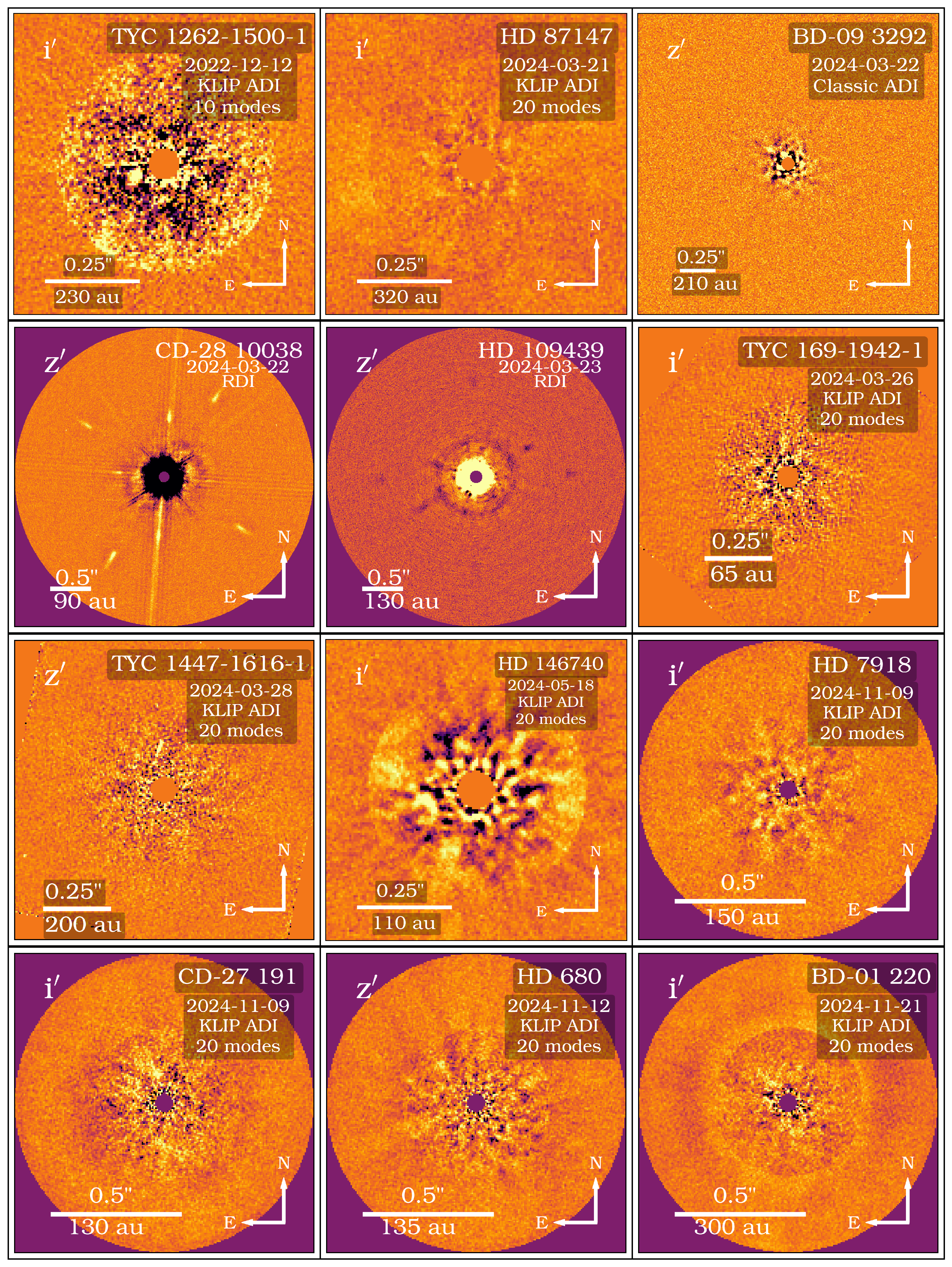}
\caption{\small{Reduced images for the systems without candidate signal.  Each system was reduced using the method indicated. All point-source like signals have been determined to be speckles.}
}
\label{fig:nondet-images}
\end{figure*}

\begin{figure*}
\centering
\includegraphics[width=0.9\textwidth]{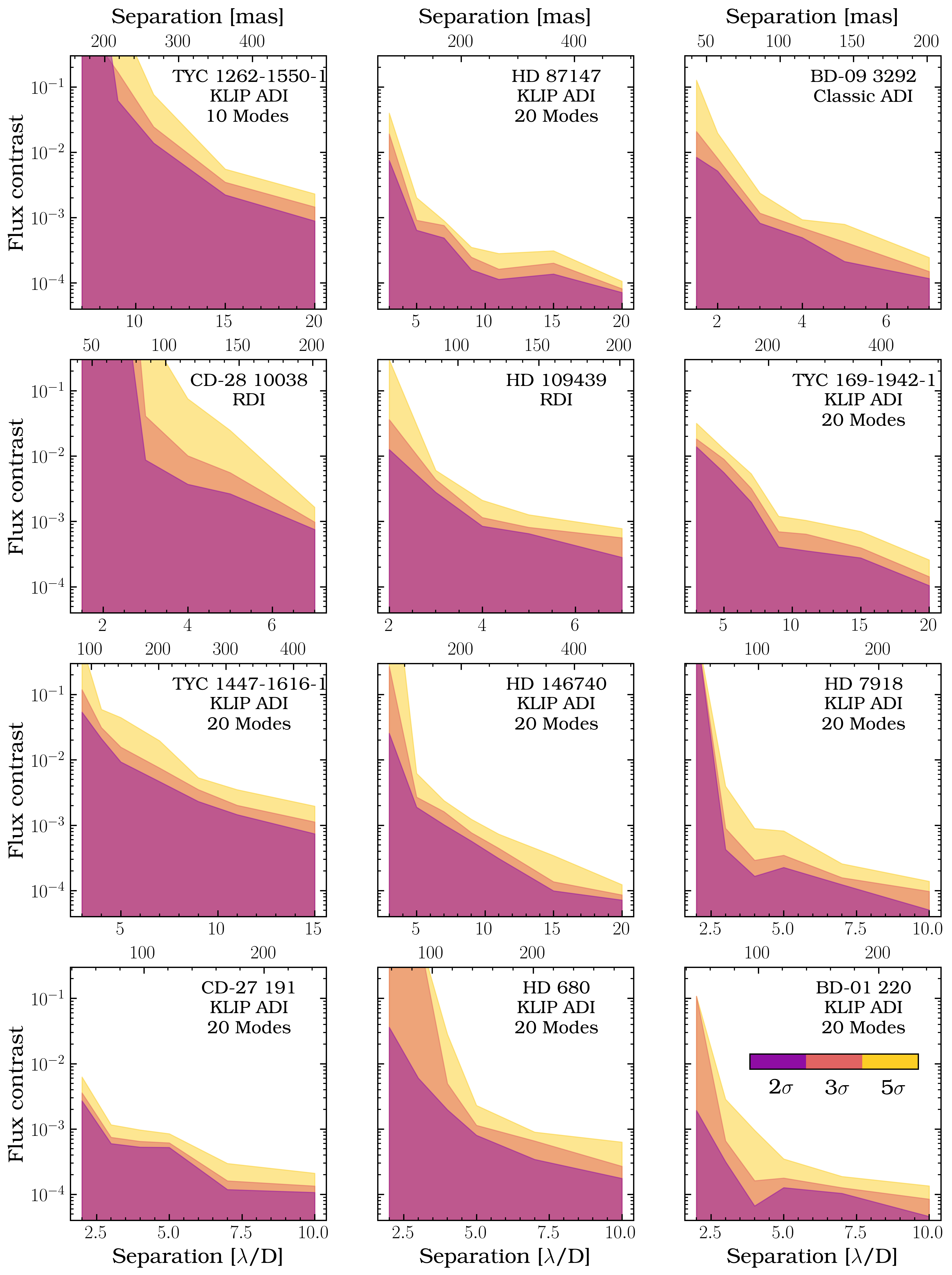}
\caption{\small{Contrast curves for non-detections.  The colored regions mark the 2-, 3-, and 5-$\sigma$ limits as a function of separation for each system and reduction method given. Separation is given in both $\lambda$/D units (bottom) and milliarcseconds (top); contrasts are given in flux contrast units.  The black line gives the 1-$\sigma$ noise floor for each observation. The classical ADI reduction got lower limits for BD-09~3292 than KLIP ADI. Contrast curves were computed using the method described in Section \ref{sec: cont-curves}
}}
\label{fig:nondet-contrastcurves}
\end{figure*}

\begin{figure*}
\centering
\includegraphics[width=0.95\textwidth]{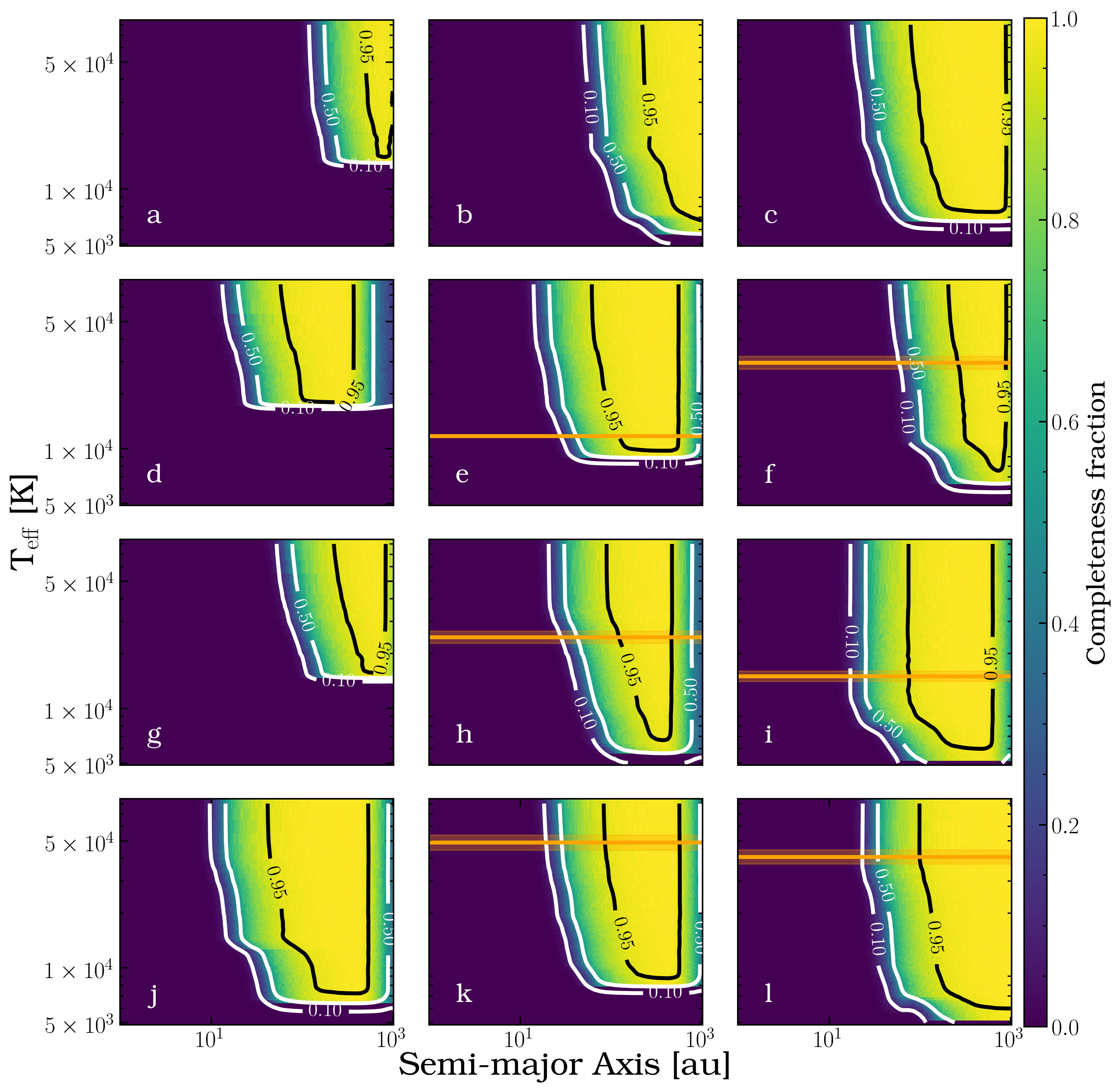}
\caption{\small{Completeness maps for non-detections. Orange horizontal lines mark the WD companion \Teff\ and uncertainty predicted in \citetalias{Ren2020AFGKSample}. a: TYC 1262-1500-1, b: HD 87147, c: BD-09 3292, d: CD-28 10038, e: HD 109439, f: TYC 169-1942-1 g: TYC 1447-1616-1, h: HD 146740, i: HD 7918, j: CD-27 191, k: HD 680, l: BD-01 220.
}}
\label{fig:nondet-composite-compl-maps}
\end{figure*}

\begin{figure*}
\centering
\includegraphics[width=0.48\textwidth]{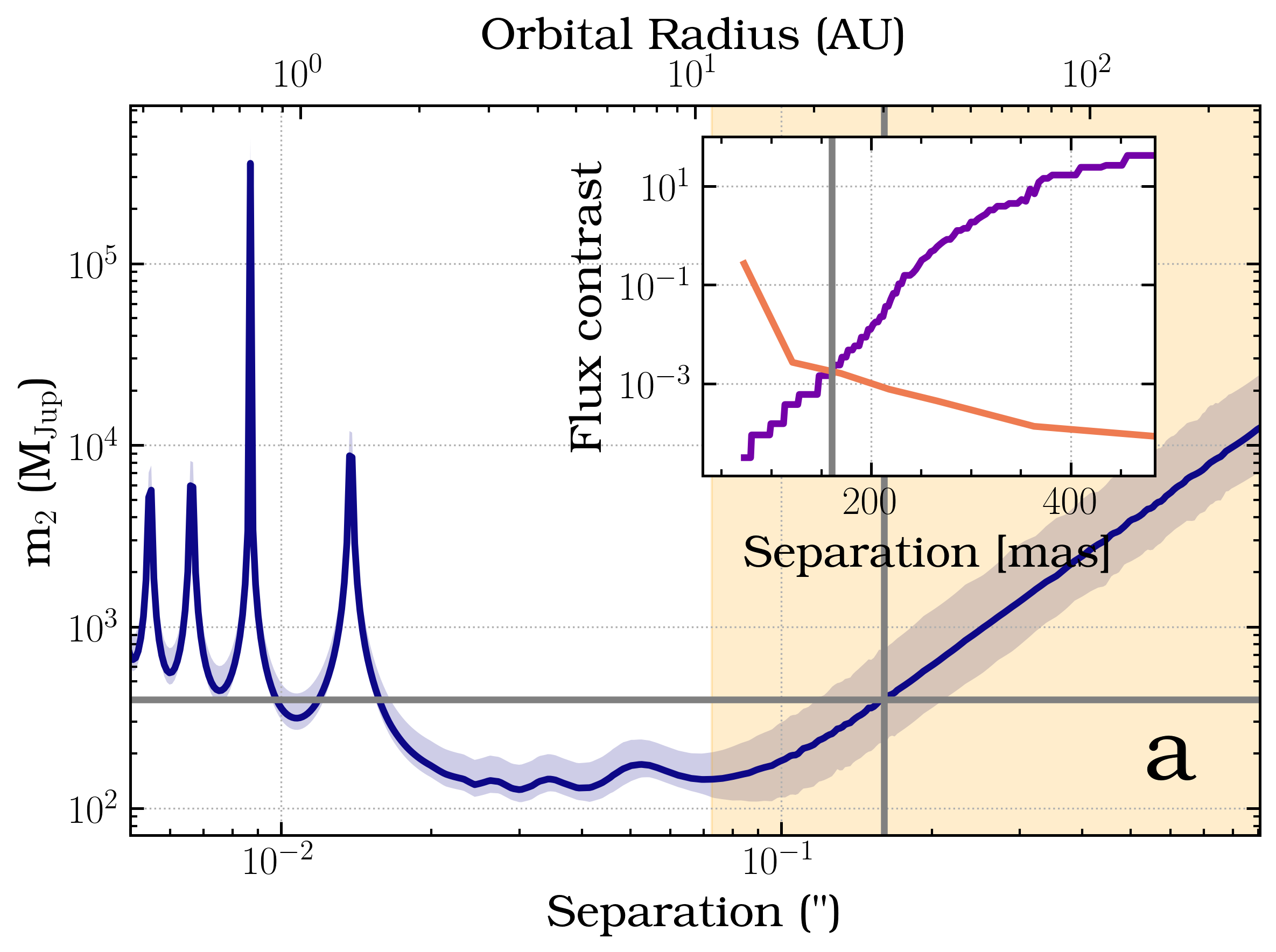}
\includegraphics[width=0.48\textwidth]{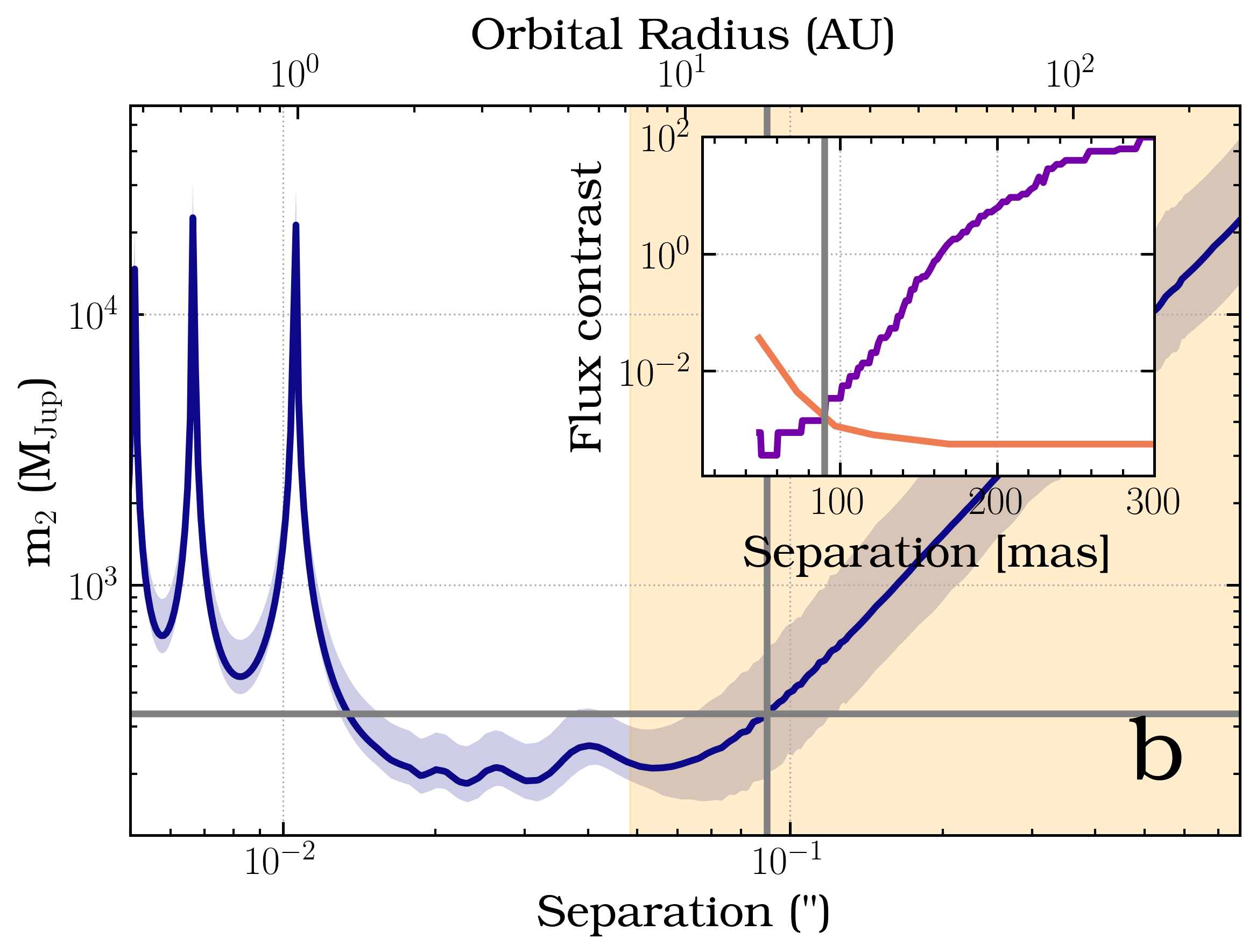}\\
\includegraphics[width=0.48\textwidth]{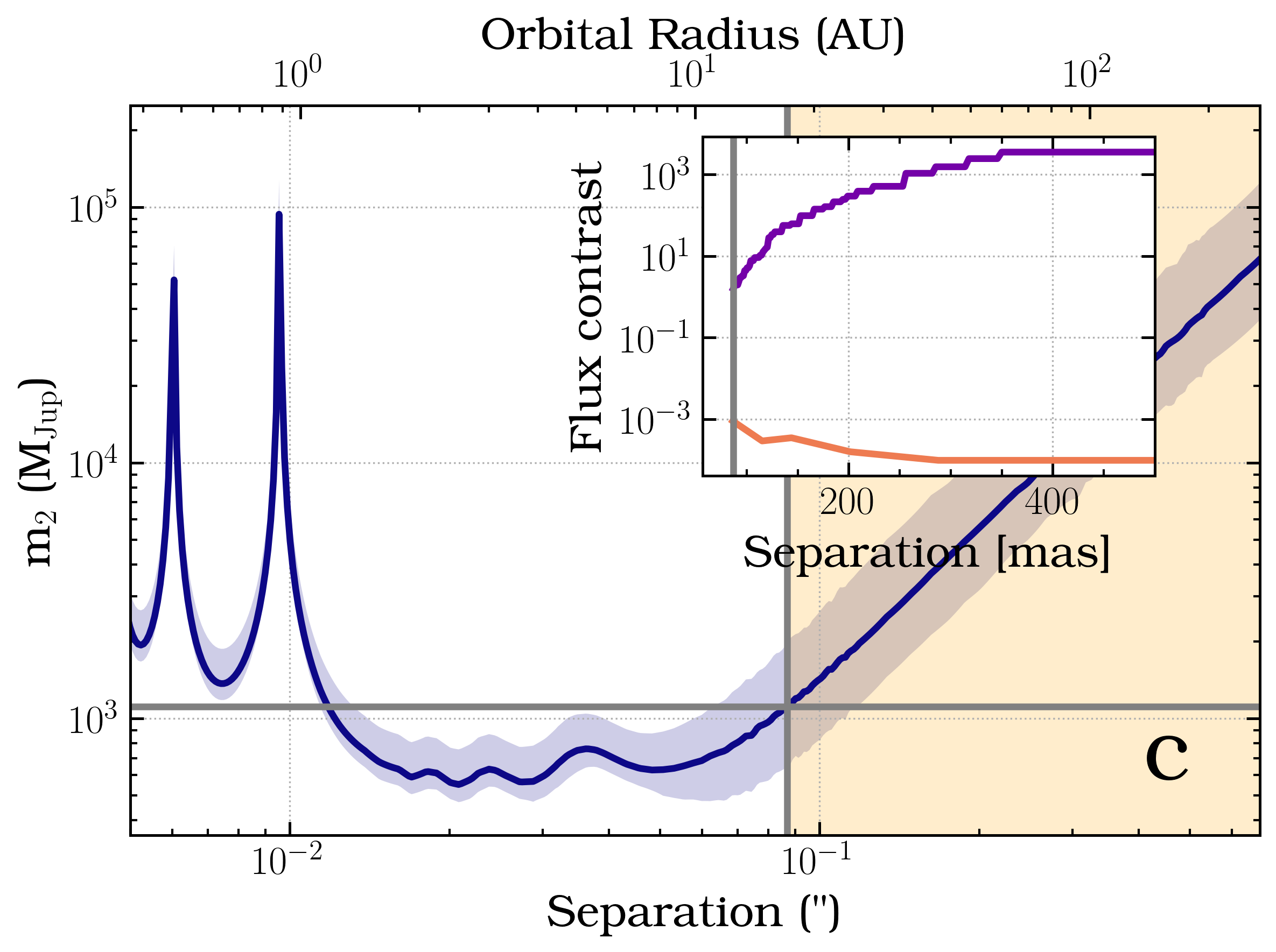}
\includegraphics[width=0.48\textwidth]{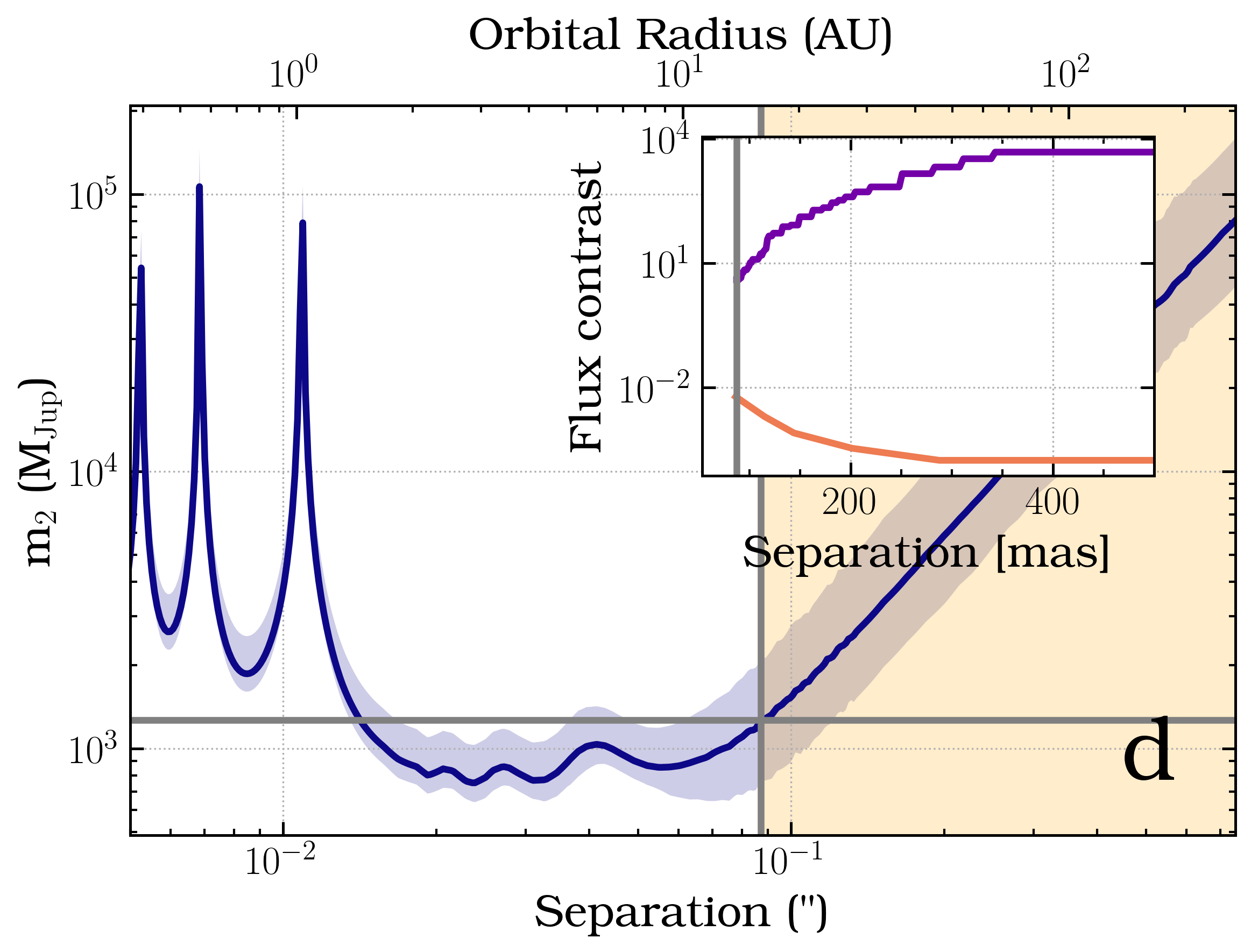}
\caption{\small{Proper motion anomaly plots for the four stars with non-detections with \textsl{Hipparcos} and \textit{Gaia} astrometry, generated using the methodology of \cite{Kervella2022}. The blue curve shows the mass of an unresolved companion that would cause the observed acceleration as a function of separation. The orange region marks where our imaging is sensitive. The inset axes shows the flux contrast corresponding to those minimum masses (purple) and our 3$\sigma$ contrast limit (red). The grey vertical lines mark where they cross, the grey horizontal line marks the minimum mass at that limit. Our contrast curves rule out at 3$\sigma$ companions within the orange region beyond the grey vertical line. a: HD~146740, b: HD~109439, c: HD~7918, d: HD~680.
}}
\label{fig:nondet-pma-plots}
\end{figure*}

\begin{deluxetable*}{cccc|ccc|ccc|cc}
\tablecaption{Observations \label{tab:observations}}
\tablehead{
\colhead{TYC} & \colhead{Obs Date} &  \colhead{$i^{\prime}$} & \colhead{Seeeing\tablenotemark{a}} & \multicolumn{3}{c}{Camsci1} & \multicolumn{3}{c}{Camsci2} & \colhead{Rotation } & \colhead{Total Int Time} \\
\colhead{} & \colhead{} &  \colhead{[mag]} &  \colhead{[arcsec]} & \colhead{Filter} & \colhead{Exp Time [s]} & \colhead{EM Gain} &  \colhead{Filter} & \colhead{Exp Time [s]} & \colhead{EM Gain} & \colhead{[deg]} & \colhead{[min]} \\
}
\startdata
\hline
    4831-473-1 & 2022-12-04 & 10.0 & 0.7 & $i^{\prime}$ & 0.5 & 5 & $r^{\prime}$ & 0.5 & 5 & 5.6 & 20.4 \\
    169-1942-1 & 2022-12-14 & 9.7 & 0.8 & $i^{\prime}$ & 1 & 5 & $r^{\prime}$ & 1 & 5 & 17.7 & 44.4 \\
    1262-1500-1 & 2022-12-14 & 10.5 & 0.7 & $i^{\prime}$ & 1 & 100 & $r^{\prime}$ & 1 & 100 & 10.0 & 34.8 \\
    4831-473-1 & 2024-03-21 & 10.0 & 0.5 & $z^{\prime}$ & 0.23 & 20 & $i^{\prime}$ & 0.23 & 50 & 11.7 & 26.4 \\
    ~ & ~ & ~ & ~ & $r^{\prime}$ & 0.23 & 50 & $g^{\prime}$ & 0.23 & 100 & 3.3 & 6.0 \\
    4913-1224-1 & 2024-03-21 & 5.9 & 0.7 & $z^{\prime}$ & 0.23 & 20 & $i^{\prime}$ & 0.23 & 1 & 0.3 & 1.8 \\ 
    ~ & ~ & ~ & ~ & $r^{\prime}$ & 0.23 & 10 & $g^{\prime}$ & 0.23 & 100 & 0.5 & 3.0 \\ 
    5480-589-1 & 2024-03-21 & 8.1 & 0.9 & $z^{\prime}$ & 0.23 & 20 & $i^{\prime}$ & 0.23 & 1 & 40.5 & 67.8 \\ 
    ~ & ~ & ~ & ~ & $r^{\prime}$ & 0.23 & 5 & $g^{\prime}$ & 0.23 & 50 & 14.7 & 18.0 \\ 
    5512-916-1 & 2024-03-22 & 8.7 & 0.6 & $z^{\prime}$ & 0.23 & 10 & $i^{\prime}$ & 0.23 & 5 & 33.8 & 51.6 \\
    ~ & ~ & ~ & ~ & $r^{\prime}$ & 0.23 & 10 & $g^{\prime}$ & 0.23 & 100 & 16.3 & 25.8 \\ 
    1385-562-1 & 2024-03-22 & 9.8 & 0.7 & $z^{\prime}$ & 0.23 & 100 & $i^{\prime}$ & 0.23 & 50 & 1.7 & 5.4 \\ 
    ~ & ~ & ~ & ~ & $r^{\prime}$ & 0.23 & 199 & $g^{\prime}$ & 0.23 & 1000 & 2.5 & 7.2 \\ 
    4865-655-1 & 2024-03-22 & 10.7 & 0.5 & $z^{\prime}$ & 0.23 & 100 & $i^{\prime}$ & 0.23 & 100 & 6.4 & 13.8 \\ 
    ~ & ~ & ~ & ~ & $r^{\prime}$ & 0.23 & 50 & $g^{\prime}$ & 0.23 & 100 & 6.8 & 14.4 \\ 
    1451-111-1 & 2024-03-22 & 10.8 & 0.5 & $z^{\prime}$ & 0.23 & 100 & $i^{\prime}$ & 0.23 & 50 & 1.9 & 6.6 \\ 
    ~ & ~ & ~ & ~ & $r^{\prime}$ & 0.23 & 100 & $g^{\prime}$ & 0.23 & 1000 & 2.6 & 8.4 \\ 
    6712-1511-1 & 2024-03-22 & 10.3 & 0.5 & $z^{\prime}$ & 0.23 & 100 & $i^{\prime}$ & 0.23 & 30 & 1.0 & 31.8 \\ 
    ~ & ~ & ~ & ~ & $r^{\prime}$ & 0.23 & 100 & $g^{\prime}$ & 0.23 & 1000 & 1.0 & 18.6 \\ 
    877-681-1 & 2024-03-23 & 7.9 & 0.7 & $z^{\prime}$ & 0.23 & 5 & $i^{\prime}$ & 0.23 & 1 & 7.1 & 28.8 \\ 
    ~ & ~ & ~ & ~ & $r^{\prime}$ & 0.23 & 5 & $g^{\prime}$ & 0.23 & 100 & 4.0 & 18.6 \\ 
    5518-135-1 & 2024-03-23 & 8.4 & 0.8 & $z^{\prime}$ & 0.23 & 10 & $i^{\prime}$ & 0.23 & 5 & 15.4 & 41.4 \\ 
    ~ & ~ & ~ & ~ & $r^{\prime}$ & 0.23 & 10 & $g^{\prime}$ & 0.23 & 200 & 1.0 & 18.6 \\ 
    169-1942-1 & 2024-03-26 & 9.7 & 0.6 & $z^{\prime}$ & 0.23 & 70 & $i^{\prime}$ & 0.23 & 30 & 2.7 & 15.6 \\ 
    ~ & ~ & ~ & ~ & $r^{\prime}$ & 0.23 & 10 & $g^{\prime}$ & 0.23 & 200 & 4.5 & 22.2 \\ 
    288-976-1 & 2024-03-27 & 9.4 & 0.6 & $z^{\prime}$ & 0.23 & 50 & $i^{\prime}$ & 0.23 & 10 & 14.5 & 64.2 \\ 
    ~ & ~ & ~ & ~ & $r^{\prime}$ & 0.23 & 50 & $g^{\prime}$ & 0.23 & 700 & 0.7 & 3.6 \\ 
    1447-1616-1 & 2024-03-28 & 10.7 & 0.6 & $z^{\prime}$ & 2 & 100 & $i^{\prime}$ & 2 & 100 & 17.1 & 49.2 \\ 
    ~ & ~ & ~ & ~ & $r^{\prime}$ & 2 & 1000 & $g^{\prime}$ & 2 & 1000 & 2.3 & 5.4 \\ 
    368-1591-1 & 2024-05-18 & 5.7 & 1.1 & $z^{\prime}$ & 0.23 & 1 & $i^{\prime}$ & 0.23 & 1 & 18.3 & 43.8 \\
    26-39-1 & 2024-11-09 & 7.5 & 0.7 & $z^{\prime}$ & 0.23 & 1 & $i^{\prime}$ & 0.23 & 1 & 18.5 & 49.8 \\ 
    6423-1892-1 & 2024-11-09 & 9.9 & 1.2 & $z^{\prime}$ & 0.23 & 200 & $i^{\prime}$ & 0.23 & 30 & 108.5 & 72.6 \\ 
    5-436-1 & 2024-11-12 & 6.9 & 1.0 & $z^{\prime}$ & 0.25 & 50 & $i^{\prime}$ & 0.25 & 40 & 19.7 & 52.8 \\ 
    4685-1113-1 & 2024-11-21 & 9.1 & 0.7 & $z^{\prime}$ & 0.25 & 30 & $i^{\prime}$ & 0.25 & 20 & 16.4 & 36.0 \\ 
\enddata
\tablenotetext{a}{Average seeing reported through the observation as reported by the on-site differential image motion monitor at 0.5$\mu$m.}
\end{deluxetable*}

\section*{Acknowledgements}

The authors wish to thank the anonymous referee for helpful comments and suggestions that improved the paper.

L.A.P.~acknowledges research support from the NSF Graduate Research Fellowship and from the University of Michigan through the ELT Fellowship Program.  This material is based upon work supported by the National Science Foundation Graduate Research Fellowship Program under Grant No. DGE-1746060. 

J.D.L.~thanks the Heising-Simons Foundation (Grant \#2020-1824) and NSF AST (\#1625441, MagAO-X).

MagAO-X was developed with support from the NSF MRI Award \#1625441. The Phase II upgrade program is made possible by the generous support of the Heising-Simons Foundation.

We thank the LCO and Magellan staffs for their outstanding assistance throughout our observing runs.

This work has made use of data from the European Space Agency (ESA) mission
\textsl{Gaia} (\url{https://www.cosmos.esa.int/gaia}), processed by the \textsl{Gaia}
Data Processing and Analysis Consortium (DPAC,
\url{https://www.cosmos.esa.int/web/gaia/dpac/consortium}). Funding for the DPAC
has been provided by national institutions, in particular the institutions
participating in the {\it Gaia} Multilateral Agreement.

This research made use of Photutils, an Astropy package for
detection and photometry of astronomical sources (Bradley et al.
2023).

This work made use of Astropy:\footnote{http://www.astropy.org} a community-developed core Python package and an ecosystem of tools and resources for astronomy \citep{astropy:2013, astropy:2018, astropy:2022}

This work has made use of the Python package GaiaXPy, developed and maintained by members of the Gaia Data Processing and Analysis Consortium (DPAC), and in particular, Coordination Unit 5 (CU5), and the Data Processing Centre located at the Institute of Astronomy, Cambridge, UK (DPCI).

\bibliography{Polluted-white-dwarfs}{}
\bibliographystyle{aasjournal}

\end{document}